\documentclass[twocolumn,english,aps,pra,superscriptaddress]{revtex4-2}

\usepackage{mathpazo}

\usepackage[T1]{fontenc}
\usepackage[latin9]{inputenc}
\usepackage{xcolor}
\definecolor{lyxbackgroundcolor}{rgb}{1, 1, 1}
\colorlet{page_backgroundcolor}{lyxbackgroundcolor}
\pagecolor{page_backgroundcolor}
\usepackage{babel}
\usepackage{amsmath}
\usepackage{amssymb}
\usepackage{graphicx}
\PassOptionsToPackage{normalem}{ulem}
\usepackage{ulem}
\usepackage[bookmarks=false,
 breaklinks=true,pdfborder={0 0 0},pdfborderstyle={},backref=false,colorlinks=true]
 {hyperref}
\hypersetup{
 pdfborderstyle=,citecolor=blue,filecolor=black,linkcolor=red,urlcolor=blue}

\makeatletter

\providecolor{lyxadded}{rgb}{0,0,1}
\providecolor{lyxdeleted}{rgb}{1,0,0}
\DeclareRobustCommand{\mklyxadded}[1]{\textcolor{lyxadded}\bgroup#1\egroup}
\DeclareRobustCommand{\mklyxdeleted}[1]{\textcolor{lyxdeleted}\bgroup\mklyxsout{#1}\egroup}
\DeclareRobustCommand{\mklyxsout}[1]{\ifx\\#1\else\sout{#1}\fi}

\makeatother

\begin{document}
\title{Saturable nonlinear Schrödinger equation with space- and time-dependent
variable coefficients}
\author{Maurilho R. da Rocha}
\affiliation{Instituto de Física, Universidade Federal de Goiás, 74.960-900, Goiânia,
Goiás, Brazil}
\author{Mateus C. P. dos Santos}
\affiliation{Instituto de Ciências Tecnológicas e Exatas, Universidade Federal
do Triângulo Mineiro, 38064-200, Uberaba, Minas Gerais, Brazil}
\affiliation{Instituto Federal do Maranhão, PPGCTM, 65030-005, São Luís, Maranhão,
Brazil}
\author{Wesley B. Cardoso}
\email{wesleybcardoso@ufg.br}

\affiliation{Instituto de Física, Universidade Federal de Goiás, 74.960-900, Goiânia,
Goiás, Brazil}
\begin{abstract}
In this paper we study the dynamics and stability of localized solutions
in a saturable nonlinear Schrödinger equation with space- and time-dependent
variable coefficients. Using a variational approach and numerical
simulations, we analyze the effects of different external potential
configurations. Our results reveal that the stability of the solutions
is highly sensitive to the modulation parameters, leading to the emergence
of alternating stable and unstable regions as a function of the modulation
frequency. These findings provide valuable insights into the control
of localized structures in nonlinear wave systems, with potential
implications for optical waveguides, Bose-Einstein condensates, and
other nonlinear media.
\end{abstract}
\maketitle

\section{Introduction}

The nonlinear Schrödinger (NLS) equation is one of the fundamental
frameworks for describing wave propagation phenomena in dispersive
and nonlinear media \citep{Sulem_04}, with applications ranging from
nonlinear optics \citep{Kivshar_03,Agrawal_13} to Bose-Einstein condensates
(BECs) \citep{Pethick_08,Pitaevskii_16}. In many physical scenarios,
the properties of the medium may vary in space and time, necessitating
the consideration of NLS versions with modulated coefficients. The
ability to control such coefficients enables the design and manipulation
of localized wave dynamics, such as solitons and optical pulses, opening
new possibilities for applications in optical communications, signal
processing, and experiments with ultracold atoms.

A significant advancement in this direction was presented by Belmonte-Beitia
\emph{et al}. \citep{Belmonte-Beitia_PRL08}, who investigated the
NLS equation with spatially and temporally dependent variable coefficients.
By employing the similarity transformation technique, the authors
demonstrated that the modulation of nonlinearity and external potential
can induce exact solitonic solutions, enabling refined control over
their dynamics. These findings motivate a broader exploration of the
effects of spatiotemporal variations in nonlinearity, particularly
in physical systems where precise manipulation of interactions is
feasible \citep{Avelar_PRE09,Avelar_PRE10,Calaca_CNSNS14,Calaca_EPJST18,Calaca_OQE17,Cardoso_BJP21,Cardoso_CNSNS17,Cardoso_NA10,Cardoso_ND21,Cardoso_PLA10,Cardoso_PLA10-2,Cardoso_PRE12,Cardoso_PRE13,Maddouri_PLA24,Miranda_OQE24,Nath_EPJD22,Oztas_PLA24,Rocha_ND23,Salasnich_PRA14,Santos_ND22,Saravanan_CNSNS19,Uthayakumar_FP20,Yan_PLA10}.
As an example, in Refs. \citep{Avelar_PRE09,Belmonte-Beitia_JPA09},
the authors analyzed more general configurations, including the presence
of spatially and temporally modulated cubic and quintic nonlinearities,
revealing a rich dynamical structure for localized solutions. These
studies indicate that different functional forms for the modulation
of nonlinearity and confinement can lead to the emergence of new wave
propagation regimes in such a system.

In another context, by considering a NLS equation with spatially and
temporally modulated quadratic and cubic nonlinearities, Ref. \citep{Cardoso_CNSNS17}
investigated exact localized solutions, demonstrating the possibility
of generating solitonic solutions that exhibit complex oscillatory
behaviors. These behaviors are highly dependent on the parameters
governing the trapping potential and the nonlinear interactions. Furthermore,
recent studies on spatially and temporally modulated saturable NLS
equations have shown the feasibility of the existence and control
of localized solutions under different physical conditions. In particular,
Ref. \citep{Rocha_ND23} analyzed the existence of stable solitons
in systems with saturable nonlinearity subject to specific modulation,
while Ref. \citep{Calaca_OQE17} investigated the impact of modulation
on the robustness of these solutions, revealing regimes where stability
can be enhanced. These results indicate that the engineering of variable
coefficients can be a powerful tool for controlling soliton propagation
in saturable media as well.

In this work, we investigate the modulation of localized solutions
in the saturable NLS equation with spatially and temporally dependent
variable coefficients. To this end, we employ the similarity transformation
technique to construct modulated solutions and subsequently test their
stability against small perturbations through direct numerical simulations.
This approach enables a detailed analysis of the robustness of the
solutions and the influence of modulation parameters on their dynamical
evolution. The present study based on the works of Refs. \citep{Calaca_OQE17,Rocha_ND23},
extending their analyses to new classes of saturable nonlinearity
modulation and exploring regimes where control over localized solutions
can be optimized. Here, differently from the previous results presented
in Refs. \citep{Calaca_OQE17,Rocha_ND23}, where the nonlinearity
was modulated solely with respect to the propagation variable, we
now incorporate modulation in both spatial and temporal coordinates.
This may render the manipulation of the equation and the derivation
of exact solutions a more challenging task compared to those obtained
previously.

The remainder of the work is structured as follows. The theoretical
model is presented in the next section, where we derive the conditions
for applying the similarity transformation. The solution to the autonomous
equation is obtained using the variational approximation described
in Sec. 3. In Sec. 4, we present the linear stability analysis. The
analytical results are detailed in Sec. 5, while the numerical simulations
are presented in Sec. 6. Finally, our conclusions and perspectives
are outlined in Sec. 7.

\section{Theoretical model \label{sec:Theoretical-model}}

Consider a nonlinear and dispersive system described by the following
NLS equation with saturable and inhomogeneous nonlinearity, expressed
in a general form as
\begin{equation}
i\psi_{t}=-\frac{1}{2}\psi_{xx}+V(x,t)\psi+\frac{g(x,t)|\psi|^{2}\psi}{1+\gamma(x,t)|\psi|^{2}},\label{main}
\end{equation}
where $\psi=\psi(x,t)$ is the field amplitude, with $\psi_{t}\equiv\frac{\partial\psi}{\partial t}$
and $\psi_{xx}\equiv\frac{\partial^{2}\psi}{\partial x^{2}}$, $V(x,t)$,
$g(x,t)$ and $\gamma(x,t)$ are parameters representing the linear,
nonlinear, and saturable terms, respectively. To solve Eq. (\ref{main}),
we employ the similarity transformation method to convert the non-autonomous
Eq. (\ref{main}) into an autonomous form. For this purpose, we start
with an \emph{ansatz} in the form
\begin{equation}
\psi=\rho(x,t)\exp[i\eta(x,t)]\Phi[\zeta(x,t)].\label{ansatz}
\end{equation}
We use this \emph{ansatz} to rewrite Eq. (\ref{main}) as
\begin{equation}
E\Phi=-\frac{1}{2}\Phi_{\zeta\zeta}+\frac{G|\Phi|^{2}\Phi}{1+\Gamma|\Phi|^{2}},\label{autonomous}
\end{equation}
where $E$ is the eigenvalue of the nonlinear equation above. The
parameters $G$ and $\Gamma$ are constants, and $\Phi=\Phi(\zeta)$
is the amplitude of the field of the autonomous equation, with the
coordinate $\zeta=\zeta(x,t)$. Substituting the \emph{ansatz} Eq.
(\ref{ansatz}) into Eq. (\ref{main}), we obtain Eq. (\ref{autonomous}),
leading to the following conditional equations:
\begin{equation}
2\rho\frac{\partial\rho}{\partial t}+\frac{\partial}{\partial x}\left[\rho^{2}\frac{\partial\eta}{\partial x}\right]=0,\label{c1}
\end{equation}
\begin{equation}
\frac{\partial\zeta}{\partial t}+\frac{\partial\eta}{\partial x}\frac{\partial\zeta}{\partial x}=0,\label{c2}
\end{equation}
\begin{equation}
\frac{\partial}{\partial x}\left[\rho^{2}\frac{\partial\zeta}{\partial x}\right]=0.\label{c3}
\end{equation}

Here, we can introduce the function $\xi(x,t)$, such that $\zeta(x,t)=F[\xi(x,t)]$.
In this case, we express $\xi(x,t)=\alpha(t)x+\beta(t)$. Using Eqs.
(\ref{c3}) and (\ref{c2}), we obtain
\begin{equation}
\rho(x,t)=[\alpha/(\partial F/\partial\xi)]^{1/2},\label{r1}
\end{equation}
\begin{equation}
\eta(x,t)=-\frac{1}{2\alpha}\frac{\partial\alpha}{\partial t}x^{2}-\frac{1}{\alpha}\frac{\partial\beta}{\partial t}x+a(t),\label{e1}
\end{equation}
respectively, where $a(t)$ is an arbitrary temporal function. Furthermore,
we can express the linear, nonlinear, and saturation parameters in
the following form
\begin{equation}
V(x,t)=\frac{1}{2\rho}\frac{\partial^{2}\rho}{\partial x^{2}}-\frac{1}{2}\left(\frac{\partial\eta}{\partial x}\right)^{2}-\frac{\partial\eta}{\partial t}-\frac{E\alpha^{4}}{\rho^{4}},\label{V}
\end{equation}
\begin{equation}
g(x,t)=G\alpha^{4}\rho^{-6},\label{g}
\end{equation}
\begin{equation}
\gamma(x,t)=\Gamma\rho^{-2}.\label{gamma}
\end{equation}

In the present work, we will assume that the nonlinear parameter is
explicitly given by
\begin{equation}
g(x,t)=\alpha^{2}\exp\left(\frac{\xi^{2}}{b^{2}}\right),\label{antigauss}
\end{equation}
where $b$ is a real parameter that controls the nonlinear behavior.
Note that the present form of the nonlinearity corresponds to a function
that increases from the center toward the periphery of the parameter
$\xi$, which, as we will see below, is proportional to the coordinate
$x$. This anti-Gaussian form of the nonlinearity has been previously
considered, e.g., in Refs. \citep{Avelar_PRE09,Borovkova_PRE11}.
Then, using Eq. (\ref{antigauss}), we obtain
\begin{equation}
\rho(x,t)=G^{1/6}\alpha^{1/3}\exp\left(\frac{-\xi^{2}}{6b^{2}}\right),\label{nrho}
\end{equation}

\begin{equation}
\gamma(x,t)=\Gamma G^{-1/3}\alpha^{-2/3}\exp\left(\frac{\xi^{2}}{3b^{2}}\right).\label{ngamma}
\end{equation}

From Eqs. (\ref{r1}) and (\ref{nrho}), we can determine the coordinate
$\zeta=-i\sqrt{3}\sqrt{\pi}b\alpha^{1/3}\mathrm{erf}(i\xi/(\sqrt{3}b))/(2G^{1/3})$,
where $\mathrm{erf}$ denotes the well-known error function. Consequently,
the potential given by Eq. (\ref{V}) can be rewritten in the form
\begin{equation}
V(x,t)=w^{2}x^{2}+f_{1}x+f_{2}-EG^{-2/3}\alpha^{8/3}\exp\left(\frac{2\xi^{2}}{3b^{2}}\right),\label{npot}
\end{equation}
where $w^{2}(t)$, $f_{1}(t)$ and $f_{2}(t)$ are time-dependent
functions, given by
\begin{equation}
w^{2}(t)=\frac{\alpha^{4}}{18b^{4}}+\frac{1}{2\alpha}\frac{\partial^{2}\alpha}{\partial t^{2}}-\frac{1}{\alpha^{2}}\left(\frac{\partial\alpha}{\partial t}\right)^{2},\label{w2}
\end{equation}
\begin{equation}
f_{1}(t)=\frac{\alpha^{3}\beta}{18b^{4}}+\frac{1}{\alpha}\frac{\partial^{2}\beta}{\partial t^{2}}-\frac{2}{\alpha^{2}}\frac{\partial\alpha}{\partial t}\frac{\partial\beta}{\partial t},\label{f1}
\end{equation}
\begin{equation}
f_{2}(t)=-\frac{\alpha^{2}}{6b^{2}}+\frac{\alpha^{2}\beta^{2}}{18b^{4}}-\frac{1}{2\alpha^{2}}\left(\frac{\partial\beta}{\partial t}\right)^{2}-\frac{\partial a}{\partial t}.\label{f2}
\end{equation}

\section{Variational Solution}

Next, to solve the autonomous NLS equation given by Eq. (\ref{autonomous}),
we employ a variational method to obtain an approximate solution.
To this end, we identify the Lagrangian density as
\begin{equation}
\mathcal{L}=-E(\Phi\Phi^{*})+\frac{1}{2}|\Phi_{\zeta}|^{2}+\frac{G}{\Gamma}|\Phi|^{2}-\frac{G}{\Gamma^{2}}\ln(1+\Gamma|\Phi|^{2}),\label{LD}
\end{equation}
which produces an equation of motion in the form given by Eq. (\ref{autonomous}).
Next, we assume a trial function with a hyperbolic profile
\begin{equation}
\Phi=A\mathrm{sech}(B\zeta),\label{VA}
\end{equation}
where $A$ and $B$ are variational parameters. Then, substituting
Eq. (\ref{VA}) into Eq. (\ref{LD}), we obtain the Lagrangian density
in the form
\begin{eqnarray}
\mathcal{L} & = & -EA^{2}\mathrm{sech}^{2}(B\zeta)+\frac{G}{\Gamma}A^{2}\mathrm{sech}^{2}(B\zeta)\nonumber \\
 & + & \frac{1}{2}A^{2}B^{2}\mathrm{sech}^{2}(B\zeta)\tanh^{2}(B\zeta)\nonumber \\
 & - & \frac{G}{\Gamma^{2}}\ln(1+\Gamma A^{2}\mathrm{sech}^{2}(B\zeta)).\label{LDA}
\end{eqnarray}

The Lagrangian is obtained by integrating the Lagrangian density above
over the entire space of the coordinate $\zeta$, that is,
\begin{equation}
L=\int^{\infty}_{-\infty}\mathcal{L}\,d\zeta.\label{L}
\end{equation}
Thus, by substituting Eq. (\ref{LDA}) into Eq. (\ref{L}) and performing
the integration over $\zeta$, we obtain the Lagrangian of the system.
However, to solve the integral involving the logarithmic term, it
is necessary to perform a Taylor series expansion for the term $\ln[1+\Gamma A^{2}\mathrm{sech}^{2}(B\zeta)]$,
evaluating for small values of $\Gamma$, i.e., considering $\Gamma A^{2}\mathrm{sech}^{2}(B\zeta)\ll1$.
Here, we carry out the third-order expansion in this term, and from
Eq. (\ref{L}), we obtain
\begin{equation}
L=\frac{A^{2}}{45B}\left[-90E+15B^{2}-16\Gamma GA^{4}+30GA^{2}\right].\label{Lok}
\end{equation}

From the Euler-Lagrange equation for Eq. (\ref{Lok}), that is, $\partial L/\partial A=\partial L/\partial B=0$,
we obtain the values
\begin{equation}
A_{\pm}=\frac{\sqrt{3}}{8}\sqrt{\frac{15}{\Gamma}\pm\frac{\sqrt{5}q}{G\Gamma}},\label{A}
\end{equation}
\begin{equation}
B_{\pm}=\frac{\sqrt{\frac{\pm9q5^{3/2}-5760E\Gamma+675G}{\Gamma}}}{2^{7/2}\sqrt{15}},\label{B}
\end{equation}
with $q\equiv\sqrt{45G^{2}-256EG\Gamma}$. Note that $B$ cannot be
complex, as this would cause the solution in Eq. (\ref{VA}) to lose
its localized nature. Therefore, we will select $G$ and $E$ such
that $B$ remains real (since $\zeta$ also cannot be complex, $G$
must be positive). For the solutions with $A_{-}$ and $B_{-}$, we
obtain $E<0$ or $E>5/32\Gamma$, with $G$ necessarily positive.
For the solutions $A_{+}$ and $B_{+}$, we have $0<E<5/32\Gamma$.
In the following sections, we examine specific choices of these parameter
values to assess the stability of the solutions.

\section{Linear Stability Analysis}

The linear stability analysis is performed for the autonomous equation
Eq. (\ref{F4}), using the solutions obtained with ($A_{-}$, $B_{-}$)
and ($A_{+}$, $B_{+}$). By applying the transformation $E\Phi=i\phi_{\tau}$,
with $\phi=e^{-iE\tau}\Phi$ in Eq. (\ref{autonomous}), we can rewrite
it in the form
\begin{equation}
i\phi_{\tau}=-\frac{1}{2}\phi_{\zeta\zeta}+\frac{G|\phi|^{2}\phi}{1+\Gamma|\phi|^{2}}.\label{LSA}
\end{equation}
The linear stability analysis is carried out for this equation, taking
into account the solutions obtained above through the variational
approximation. In this context, we consider the perturbed solution
given by
\begin{eqnarray}
\phi(\zeta,\tau) & = & \{\Phi(\zeta)+[v(\zeta)+w(\zeta)]e^{\lambda\tau}\nonumber \\
 & + & [v^{*}(\zeta)-w^{*}(\zeta)]e^{\lambda^{*}\tau})e^{-iE\tau},\label{solpert}
\end{eqnarray}
where $v(\zeta)$ and $w(\zeta)$ are small perturbations, i.e., $v(\zeta),w(\zeta)\ll1$,
and $\lambda$ is the corresponding eigenvalue. Substituting this
solution into Eq. (\ref{F4}) and performing a linearization process,
we obtain an eigenvalue problem in the form
\begin{equation}
L\Psi=\lambda\Psi,\label{eigen}
\end{equation}
where
\begin{equation}
L=i\begin{pmatrix}G_{0} & \frac{1}{2}\nabla^{2}+G_{1}\\
\frac{1}{2}\nabla^{2}+G_{2} & -G_{0}
\end{pmatrix},\label{opL}
\end{equation}
with
\begin{equation}
\Psi=\begin{pmatrix}v\\
w
\end{pmatrix},
\end{equation}

\begin{equation}
G_{0}=-\frac{1}{2}(\varphi^{2}-(\varphi^{*})^{2})\frac{G}{(1+\Gamma|\phi|^{2})^{2}},
\end{equation}

\begin{eqnarray}
G_{1} & = & E-\frac{G|\varphi|^{2}}{1+\Gamma|\phi|^{2}}\nonumber \\
 & - & \left[|\varphi|^{2}-\frac{1}{2}(\varphi^{2}+(\varphi^{*})^{2})\right]\frac{G}{(1+\Gamma|\phi|^{2})^{2}},
\end{eqnarray}
and

\begin{eqnarray}
G_{2} & = & E-\frac{G|\varphi|^{2}}{1+\Gamma|\phi|^{2}}\nonumber \\
 & - & \left[|\varphi|^{2}+\frac{1}{2}(\varphi^{2}+(\varphi^{*})^{2})\right]\frac{G}{(1+\Gamma|\phi|^{2})^{2}}.
\end{eqnarray}

Next, we numerically solve the eigenvalue problem (\ref{eigen}) using
a Fourier collocation method (for more details, see Ref. \citep{Yang_10}).

\begin{figure}[tb]
\centering \includegraphics[width=0.49\columnwidth]{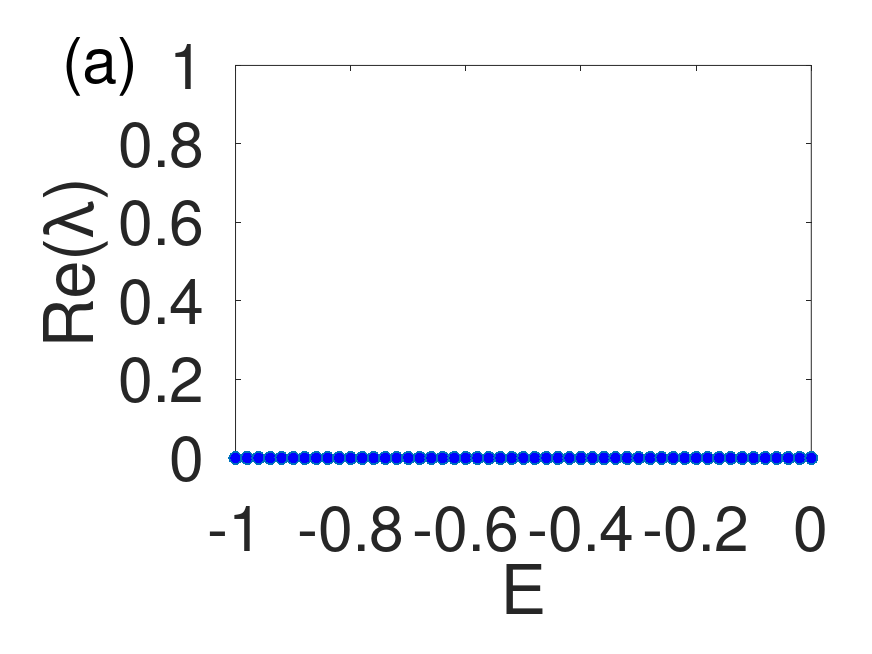} \hfil
\includegraphics[width=0.5\columnwidth]{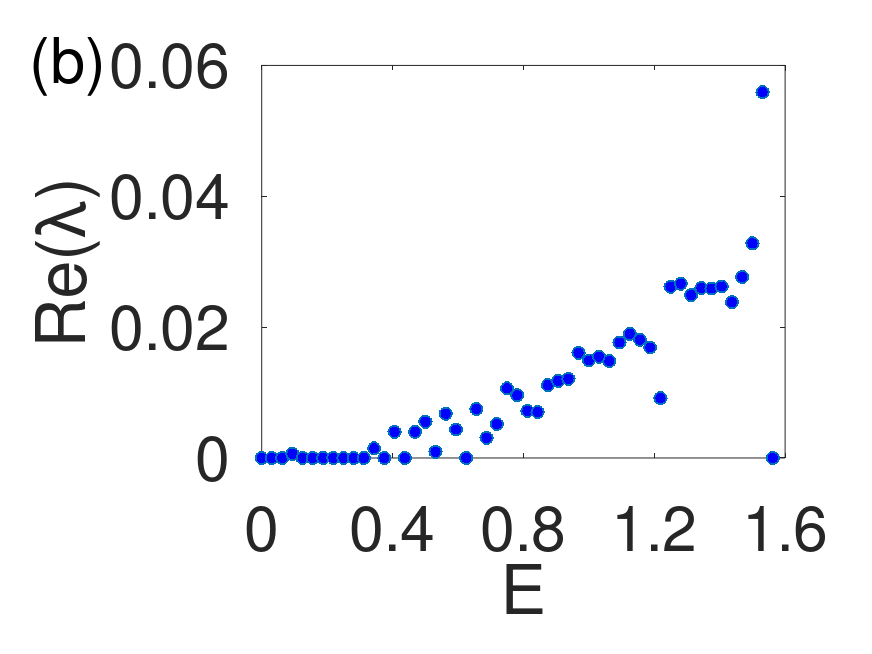} \caption{Real part of the eigenvalue $\lambda$ as a function of $E$, obtained
numerically from the variational solution (\ref{VA}), considering
the parameters (a) ($A_{-}$,$B_{-}$) and (b) ($A_{+}$,$B_{+}$)
from Eqs. (\ref{A}) and (\ref{B}), respectively. The other parameter
values used here were: $G=1.0$, $\Gamma=0.1$, and $b=2.0$.}
\label{F1}
\end{figure}


First, as an example, we choose the solution obtained with $A_{-}$
and $B_{-}$ from Eqs. (\ref{A}) and (\ref{B}), considering the
parameter values $G=1.0$, $\Gamma=0.1$ and $b=2.0$. In Fig. \ref{F1}(a),
the results obtained from numerical simulations for different values
of $E$ are shown. Note that for the range $E=[-1,0]$, the real part
of the eigenvalue of Eq. (\ref{eigen}) is always zero, implying that
the solutions obtained for these parameter values are always stable.

Next, we analyze another example in which the solution is obtained
by choosing the coefficients $A_{+}$ and $B_{+}$ in Eqs. (\ref{A})
and (\ref{B}). The results for the real part of the eigenvalue $\lambda$
are shown in Fig. \ref{F1}(b), where we use the same parameter values
as in the example from Fig. \ref{F1}(a), namely $G=1.0$, $\Gamma=0.1$
and $b=2.0$. Note that the stability region, associated with the
zero values of $\textrm{Re}(\lambda)$, is now restricted to a smaller
range of $E$ values.

We emphasize that the stability tests presented in the examples shown
in Fig. \ref{F1} are related to the solution of the autonomous equation
given by Eq. (\ref{autonomous}), which represents only a part of
the \emph{ansatz} we are constructing (\ref{ansatz}). In fact, the
physical system under consideration is described by the non-autonomous
equation (\ref{main}), and it is from this equation that we can determine
whether the solution is truly stable or unstable. However, the linear
stability analysis of the variational solution obtained for the autonomous
equation can serve as a reference for constructing modulated solutions
that also exhibit stability in the non-autonomous equation.

In the next section, we present the analytical solutions obtained
using the similarity transformation technique and study their evolution,
testing them against small perturbations. In this case, we will use
specific parameter values that favor the stability of the solutions
of the autonomous equation.

\section{Analytical Results}

In this section, we present several examples related to the choice
of coefficients $\alpha(t)$, $\beta(t)$ and $a(t)$ to modulate
the linear potential $V(x,t)$, the nonlinear coefficient $g(x,t)$,
and the saturation parameter $\gamma(x,t)$. In this context, we will
examine four different modulation patterns, all considering the parameter
values $G=1.0$, $\Gamma=0.1$, and $b=2.0$.

\textbf{Static Potential} -- First, we analyze the case of a static
potential. To achieve this, we set $\alpha(t)=1$, $\beta(t)=0$ and
$a(t)=0$ in Eqs. (\ref{w2})-(\ref{f2}). Consequently, we obtain
$w^{2}(t)=1/(18b^{4})$, $f_{1}=0$ and $f_{2}=-1/(6b^{2})$. As a
result, the amplitude can be expressed in the Gaussian form as $\rho(x,t)=G^{1/6}\alpha^{1/3}e^{-x^{2}/6b^{2}}$,
the phase velocity as $\eta=a(t)$, and the linear, nonlinear, and
saturation parameters are given by
\begin{equation}
V(x,t)=w^{2}(t)x^{2}+f_{2}(t)-EG^{-2/3}\alpha^{8/3}e^{2(\alpha x)^{2}/3b^{2}},\label{static}
\end{equation}
\begin{equation}
g(x,t)=\alpha^{2}e^{(\alpha x)^{2}/b^{2}},
\end{equation}
\begin{equation}
\gamma(x,t)=\Gamma G^{-1/3}\alpha^{-2/3}e^{x^{2}/3b^{2}}.
\end{equation}

\begin{figure}[tb]
\centering \includegraphics[width=0.47\columnwidth]{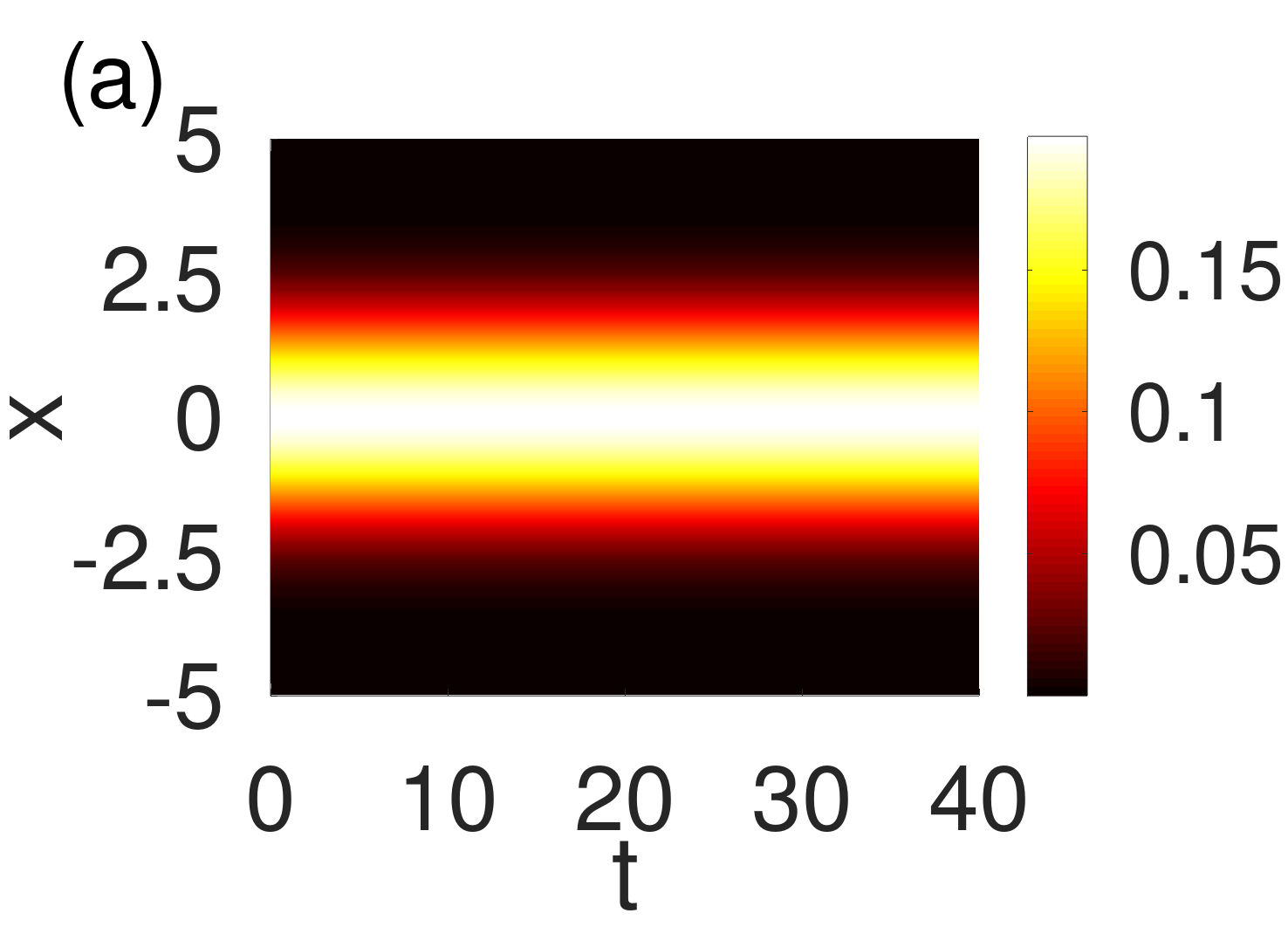} \hfil\includegraphics[width=0.47\columnwidth]{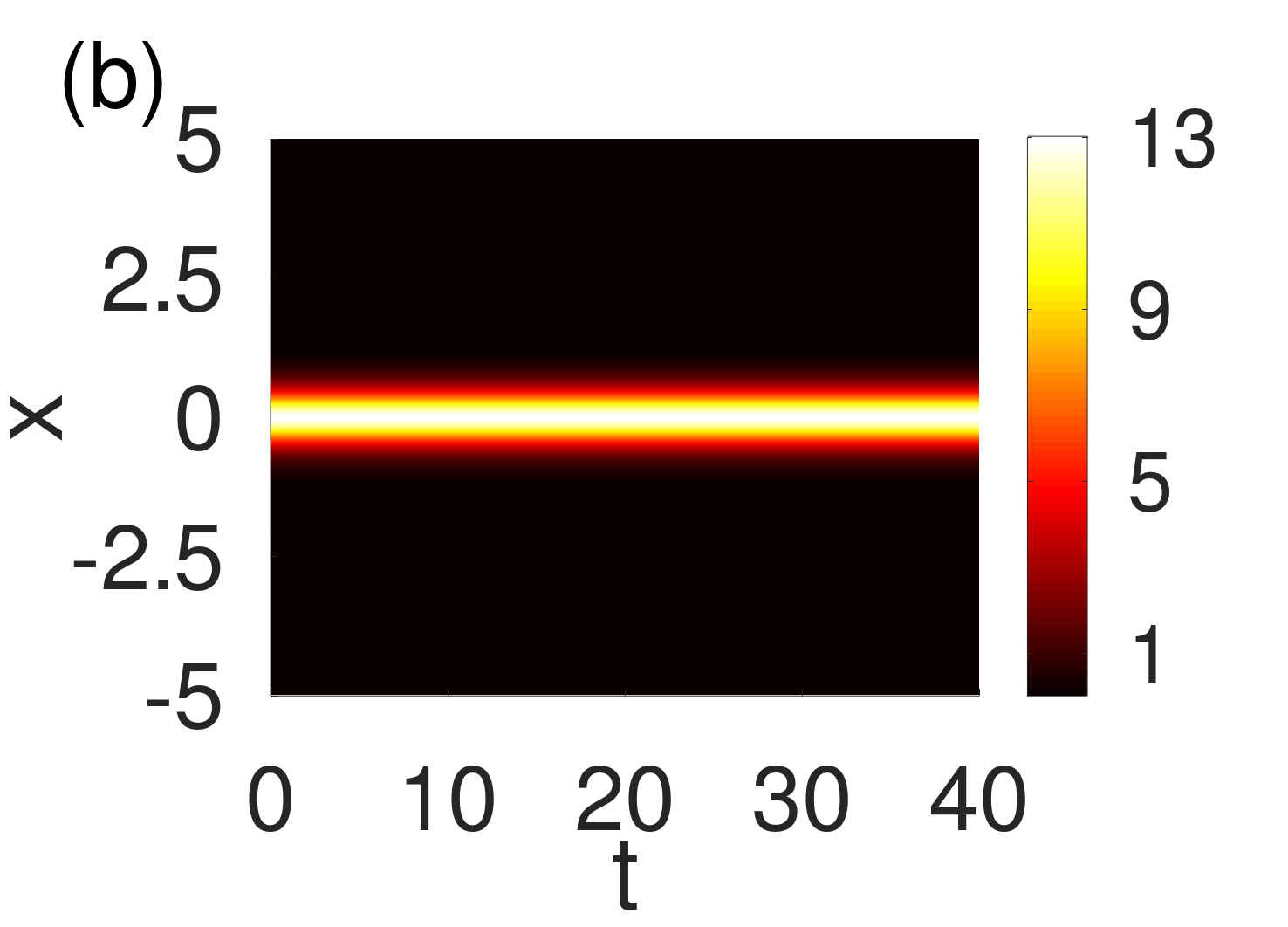}
\caption{Solution profile $|\psi|^{2}$ as a function of the coordinates $x$
and $t$ for the case with a static potential, using the variational
solutions obtained with the coefficient choices: (a) ($A_{-}$, $B_{-}$)
with $E=-0.1$ and (b) ($A_{+}$, $B_{+}$) with $E=0.4$. The other
parameter values used were $G=1.0$, $\Gamma=0.1$, and $b=2.0$.}
\label{F2}
\end{figure}

Using the results above, we plot in Fig. \ref{F2} the solution profile
$|\psi|^{2}$ as a function of the coordinates $x$ and $t$ for the
static potential. Note that the solution exhibits a stationary behavior
in time. Furthermore, the choice of different coefficients obtained
from the variational approximation, namely ($A_{-}$, $B_{-}$) or
($A_{+}$, $B_{+}$), leads to significant differences in the solution
profiles, as can be observed by comparing the results in Figs. \ref{F2}(a)
e \ref{F2}(b).

\textbf{Seesaw Potential} -- We now analyze the case of a potential
with a linear dependence on $x$ and an oscillatory behavior in $t$,
resembling a seesaw-like motion. To achieve this, we set $\alpha(t)=1$,
$\beta(t)=-\sin(w_{0}t)$, and $a(t)=0$. Consequently, we obtain
the functions $w^{2}(t)=\frac{1}{18b^{4}}$, $f_{1}(t)=-\frac{\sin(w_{0}t)}{18b^{4}}+w^{2}_{0}\sin(w_{0}t)$,
$f_{2}(t)=-\frac{1}{6b^{2}}+\frac{\sin^{2}(w_{0}t)}{18b^{4}}-\frac{w^{2}_{0}\cos^{2}(w_{0}t)}{2}$.
Moreover, the amplitude and phase of the ansatz take the forms $\rho(x,t)=G^{1/6}e^{-(x-\sin(w_{0}t))^{2}/(6b^{2})}$
and $\eta(x,t)=w_{0}\cos(w_{0}t)x+a(t)$, respectively, while the
linear, nonlinear, and saturation parameters of the system are given
by
\begin{eqnarray}
V(x,t) & = & w^{2}(t)x^{2}+f_{1}(t)x+f_{2}(t)\nonumber \\
 & - & EG^{-2/3}e^{2(x-\sin(w_{0}t))^{2}/3b^{2}},\label{linear}
\end{eqnarray}
\begin{equation}
g(x,t)=e^{(x-\sin(w_{0}t))^{2}/b^{2}},
\end{equation}
\begin{equation}
\gamma(x,t)=\Gamma G^{-1/3}e^{(x-\sin(w_{0}t))^{2}/3b^{2}}.
\end{equation}

\begin{figure}[tb]
\includegraphics[width=0.48\columnwidth]{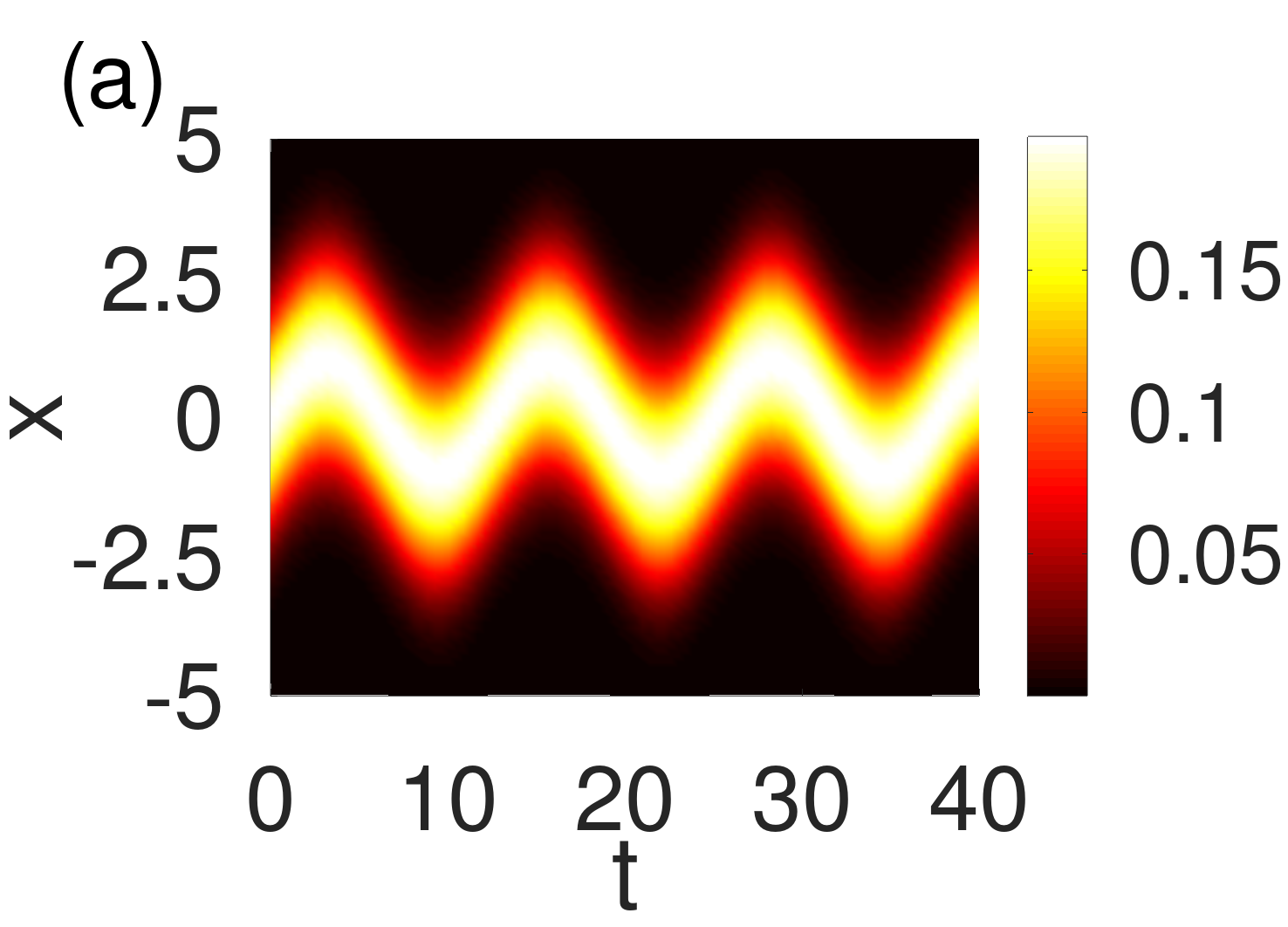} \hfil \includegraphics[width=0.48\columnwidth]{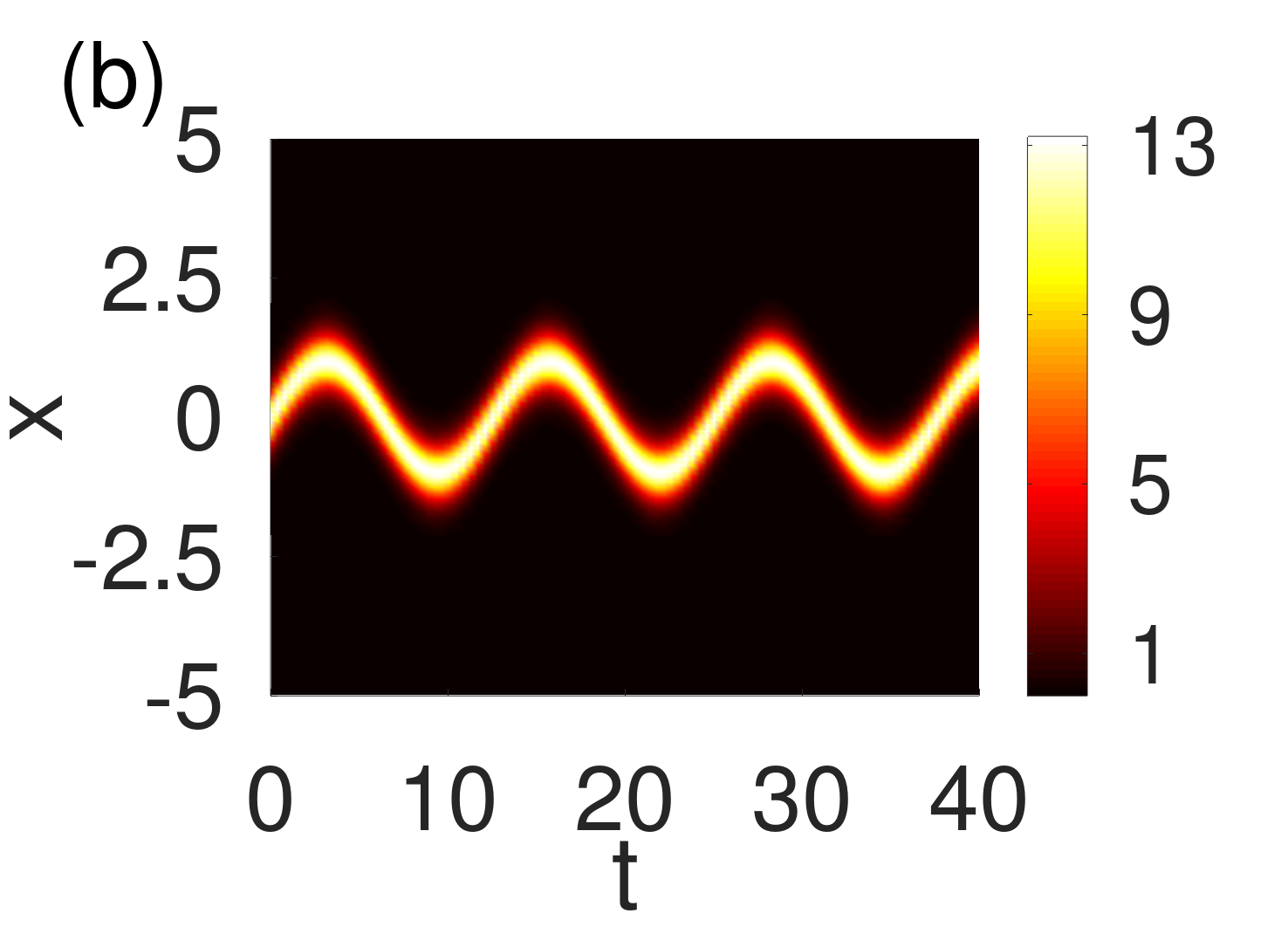}
\caption{Solution profile $|\psi|^{2}$ as a function of the coordinates $x$
and $t$ for the case with a seesaw potential, using the variational
solutions obtained with the coefficient choices: (a) ($A_{-}$, $B_{-}$)
with $E=-0.1$ and (b) ($A_{+}$, $B_{+}$) with $E=0.4$. The other
parameter values used were $G=1.0$, $\Gamma=0.1$, $b=2.0$, and
$w_{0}=0.5$.}
\label{F3}
\end{figure}

The analytical solution corresponding to this case is presented in
Fig. \ref{F3}, where it is observed that the solution profile $|\psi|^{2}$
exhibits an oscillatory motion in the center-of-mass position, generating
a seesaw-like movement for both choices of the variational parameters,
($A_{-}$, $B_{-}$) and ($A_{+}$, $B_{+}$). For the examples shown
in Fig. \ref{F3}, we used the parameter values $G=1.0$, $\Gamma=0.1$,
$b=2.0$ and $w_{0}=0.5$. Additionally, as previously observed in
Fig. \ref{F1}, the selection of variational parameters alters the
solution profile. However, the modulation follows the same pattern
for both solutions.

\textbf{Flying-Bird Potential} -- Next, we analyze the system in
the presence of a potential with a quadratic dependence on $x$ and
an oscillatory behavior in $t$. In this case, the potential undergoes
a temporal oscillation, transitioning between confining and repulsive
regimes, exhibiting a pattern reminiscent of a wing-flapping motion.
This potential is obtained by choosing $\alpha(t)=1+\sigma\sin(w_{0}t)$,
$\beta(t)=0$, and $a(t)=0$, such that
\begin{eqnarray}
w^{2}(t) & = & \frac{1}{18b^{4}}(1+\sigma\sin(w_{0}t))^{4}-\frac{\sigma w^{2}_{0}\sin(w_{0}t)}{2(1+\sigma\sin(w_{0}t))}\nonumber \\
 & - & \frac{\sigma^{2}w^{2}_{0}\cos^{2}(w_{0}t)}{(1+\sigma\sin(w_{0}t))^{2}}.
\end{eqnarray}
Furthermore, we obtain $f_{1}(t)=0$ and $f_{2}(t)=-(1+\sigma\sin(w_{0}t))^{2}/6b^{2}$.
Consequently, the amplitude and phase of the solution can be rewritten
as $\rho(x,t)=G^{1/6}[1+\sigma\sin(w_{0}t)]^{1/3}e^{-[1+\sigma\sin(w_{0}t)]^{2}x^{2}/(6b^{2})}$
and $\eta(x,t)=-\frac{\sigma w_{0}\cos(w_{0}t)}{2[1+\sigma\sin(w_{0}t)]}x^{2}$,
respectively. Finally, the linear, nonlinear, and saturation coefficients
are given by
\begin{eqnarray}
V(x,t) & = & w^{2}(t)x^{2}+f_{2}(t)\nonumber \\
 & - & EG^{-2/3}(1+\sigma\sin(w_{0}t))^{8/3}e^{2((1+\sigma\sin(w_{0}t))x)^{2}/3b^{2}},\label{fly}
\end{eqnarray}
\begin{equation}
g(x,t)=(1+\sigma\sin(w_{0}t))^{2}e^{((1+\sigma\sin(w_{0}t))x)^{2}/b^{2}},
\end{equation}
\begin{equation}
\gamma(x,t)=\Gamma G^{-1/3}(1+\sigma\sin(w_{0}t))^{-2/3}e^{((1+\sigma\sin(w_{0}t))x)^{2}/(3b^{2})}.
\end{equation}

\begin{figure}[tb]
\centering \includegraphics[width=0.48\columnwidth]{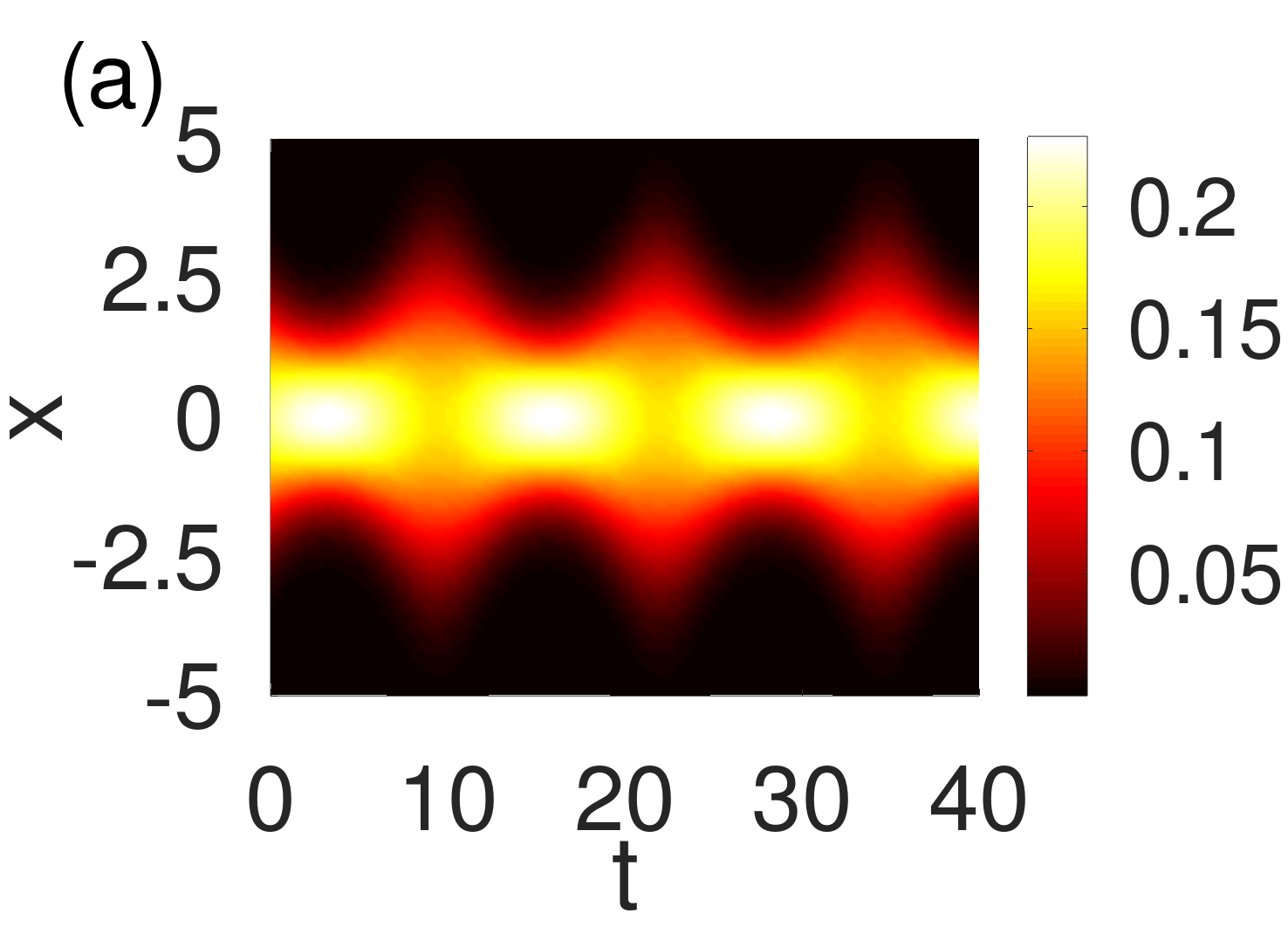} \hfil
\includegraphics[width=0.48\columnwidth]{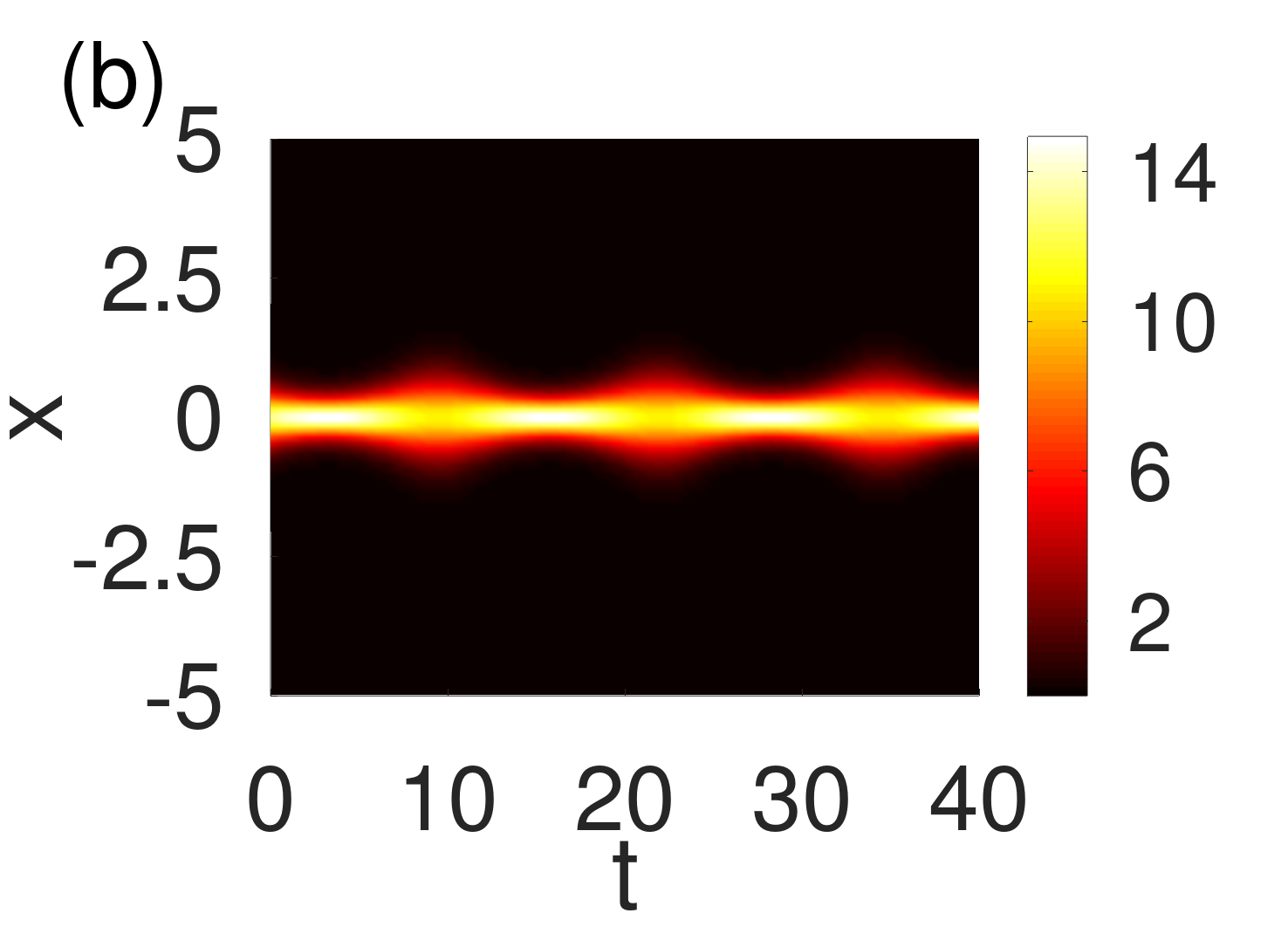} \caption{Solution profile $|\psi|^{2}$ as a function of the coordinates $x$
and $t$ for the case with a \emph{flying-bird} potential (\ref{fly}),
using the variational solutions obtained with the coefficient choices:
(a) ($A_{-}$, $B_{-}$) with $E=-0.1$ and (b) ($A_{+}$, $B_{+}$)
with $E=0.4$. The other parameter values used were $G=1.0$, $\Gamma=0.1$,
$b=2.0$, $\sigma=0.25$, and $w_{0}=0.5$.}
\label{F4}
\end{figure}

In Fig. \ref{F4}, we present the profile of the solution $|\psi|^{2}$
as a function of the coordinates $x$ and $t$. Note that it exhibits
periodic oscillations in its amplitude as it propagates in $t$. However,
we emphasize that, unlike the previous case, the center of mass of
the solution remains fixed at $x=0$. For the examples shown in Fig.
\ref{F4}, we used the parameter values $G=1.0$, $\Gamma=0.1$, $b=2.0$,
$\sigma=0.25$, and $w_{0}=0.5$.

\textbf{Mixed Potential} -- As a final example, we analyze a model
in which the potential is given by the sum of the previous two potentials,
namely the \emph{seesaw} and \emph{flying-bird} potentials, which
we refer to as the mixed potential. To construct this, we set $\alpha(t)=1+\sigma\sin(w_{0}t)$,
$\beta(t)=-\sin(w_{0}t)$, and $a(t)=0$. Consequently, we obtain
\begin{eqnarray}
w^{2}(t) & = & \frac{1}{18b^{4}}(1+\sigma\sin(w_{0}t))^{4}-\frac{\sigma w^{2}_{0}\sin(w_{0}t)}{2(1+\sigma\sin(w_{0}t))}\nonumber \\
 & - & \frac{\sigma^{2}w^{2}_{0}\cos^{2}(w_{0}t)}{(1+\sigma\sin(w_{0}t))^{2}},
\end{eqnarray}
\begin{eqnarray}
f_{1}(t) & = & -\frac{(1+\sigma\sin(w_{0}t))^{3}\sin(w_{0}t)}{18b^{4}}+\frac{w^{2}_{0}\sin(w_{0}t)}{(1+\sigma\sin(w_{0}t))}\nonumber \\
 & + & \frac{2w^{2}_{0}\sigma\cos^{2}(w_{0}t)}{(1+\sigma\sin(w_{0}t))^{2}},
\end{eqnarray}
\begin{eqnarray}
f_{2}(t) & = & -\frac{(1+\sigma\sin(w_{0}t))^{2}}{6b^{2}}-\frac{w^{2}_{0}\cos^{2}(w_{0}t)}{2(1+\sigma\sin(w_{0}t))^{2}}\nonumber \\
 & + & \frac{(1+\sigma\sin(w_{0}t))^{2}\sin^{2}(w_{0}t)}{18b^{4}}.
\end{eqnarray}

Furthermore, we can rewrite the linear, nonlinear, and saturation
coefficients as follows
\begin{eqnarray}
 &  & V(x,t)=w^{2}(t)x^{2}+f_{1}(t)x+f_{2}(t)\nonumber \\
 &  & -EG^{-2/3}(1+\sigma\sin(w_{0}t))^{8/3}e^{2(\alpha x-\sin(w_{0}t))^{2}/3b^{2}},\label{mixed}
\end{eqnarray}
\begin{equation}
g(x,t)=(1+\sigma\sin(w_{0}t))^{2}e^{((1+\sigma\sin(w_{0}t))x-\sin(w_{0}t))^{2}/b^{2}},
\end{equation}
\begin{equation}
\gamma(x,t)=\Gamma G^{-1/3}(1+\sigma\sin(w_{0}t))^{-2/3}e^{(\alpha x-\sin(w_{0}t))^{2}/(3b^{2})}.
\end{equation}

\begin{figure}[tb]
\centering \includegraphics[width=0.48\columnwidth]{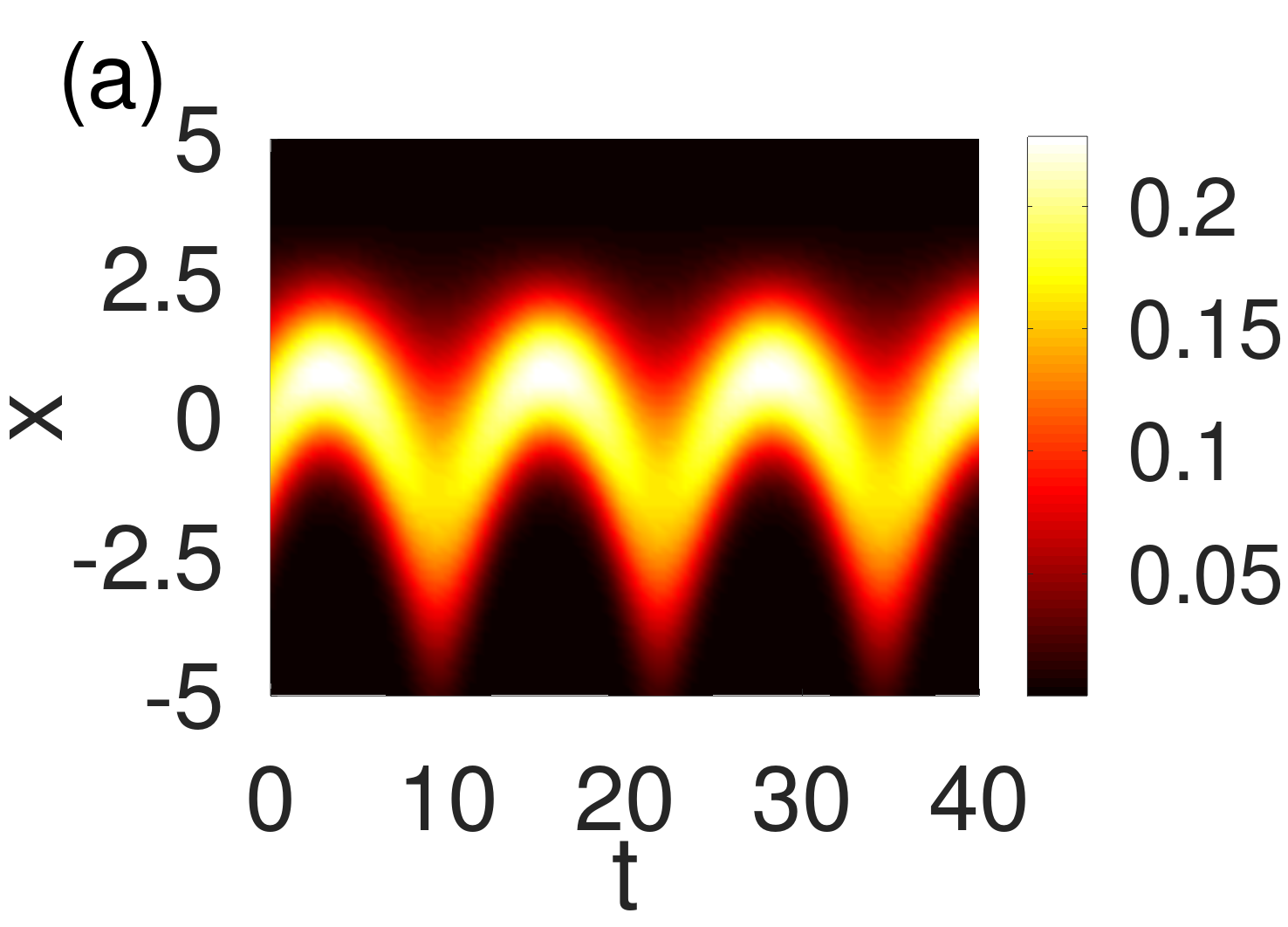} \hfil
\includegraphics[width=0.48\columnwidth]{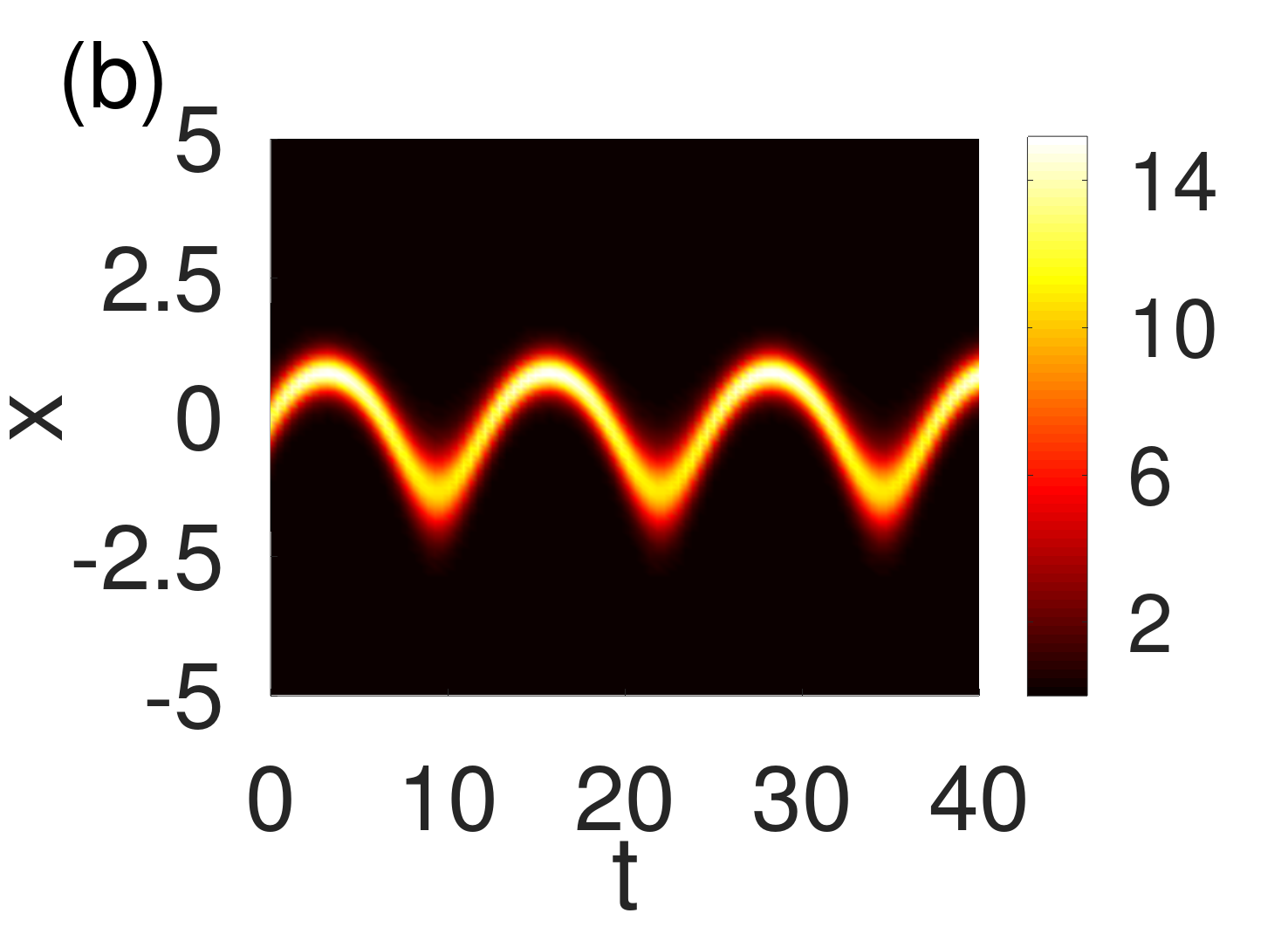} \caption{Solution profile $|\psi|^{2}$ as a function of the coordinates $x$
and $t$ for the case with a mixed potential (\ref{mixed}), using
the variational solutions obtained with the coefficient choices: (a)
($A_{-}$, $B_{-}$) with $E=-0.1$ and (b) ($A_{+}$, $B_{+}$) with
$E=0.4$. The other parameter values used were $G=1.0$, $\Gamma=0.1$,
$b=2.0$, $\sigma=0.25$, and $w_{0}=0.5$.}
\label{F5}
\end{figure}

The profile of the localized solution $|\psi|^{2}$ is presented in
Fig. \ref{F5} for two distinct variational solutions of the autonomous
equation. In this case, we observe that the oscillatory pattern of
the system\textquoteright s center of mass, previously seen in the
seesaw potential case, is also present here. However, the oscillations
occur asymmetrically in $x$. Another peculiar characteristic is the
amplitude oscillation, highlighted by the lighter regions in Fig.
\ref{F5}, which occurs only when the solution reaches positive values
of $x$. Here, we used the parameter values $G=1.0$, $\Gamma=0.1$,
$b=2.0$, $\sigma=0.25$, and $w_{0}=0.5$ to construct the plots
shown in Fig. \ref{F5}.

\section{Direct Numerical Simulations}

We now analyze the stability of the modulated solutions through direct
numerical simulations of the non-autonomous equation, Eq. (\ref{main}).
As an example, we consider the solutions obtained with the variational
parameter pair $A_{-}$ and $B_{-}$, as they demonstrated robustness
in the linear stability criteria. To this end, we examine the four
types of modulation studied analytically in the previous section to
determine whether the solutions remain stable under small perturbations
($5\%$ perturbation in the solution amplitude induced by random noise).
For the direct numerical simulations of the non-autonomous equation,
we employ the second-order split-step method. The spatial part is
solved using a spectral method. For further details, see Ref. \citep{Yang_10}.

To analyze the numerical results, we use the maximum amplitude ($h_{\max}$)
and the mean width 
\begin{equation}
\langle x^{2}\rangle=\frac{\int x^{2}|\psi|^{2}dx}{\int|\psi|^{2}dx}\label{xm}
\end{equation}
 of the solutions. Additionally, for our analysis, we select solutions
whose eigenvalue of the autonomous equation is given by $E=-0.1$.
Indeed, as previously observed, any value within the range $-1\leq E<0$
ensures the stability of the autonomous solution.

\begin{figure}[tb]
\centering \includegraphics[width=0.48\columnwidth]{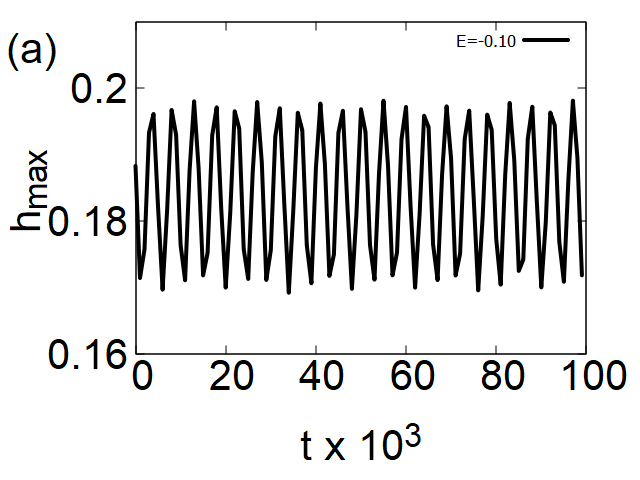} \hfil
\includegraphics[width=0.48\columnwidth]{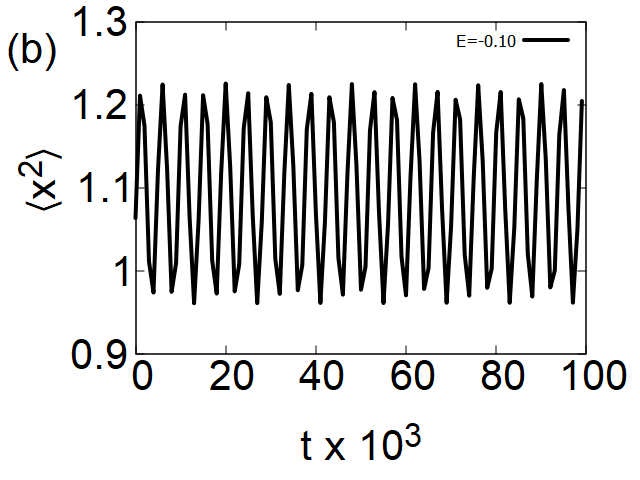} \caption{(a) Maximum amplitude of the perturbed numerical solution $|\psi|^{2}$
as a function of the coordinate $t$ for the case with a static potential
(\ref{static}), using the solutions constructed with the variational
parameters $A_{-}$ and $B_{-}$. (b) Average width of the numerical
solution as a function of $t$. The other parameter values used were
$E=-0.1$, $G=1$, $\Gamma=0.1$, and $b=2.0$.}
\label{F6}
\end{figure}

First, we consider the static potential, given by Eq. (\ref{static}).
In Fig. \ref{F6}, we present the results obtained from the simulations
for $h_{\max}$ and $\langle x^{2}\rangle$. Here, we observe that
the evolution of the perturbed solution up to $t=100000$ exhibits
a small oscillation around a fixed value, both for the amplitude and
for $\langle x^{2}\rangle$. Thus, we conclude that the modulation
produced by the static potential, considering the parameter values
under consideration, results in a stable solution.

\begin{figure}[tb]
\centering \includegraphics[width=0.48\columnwidth]{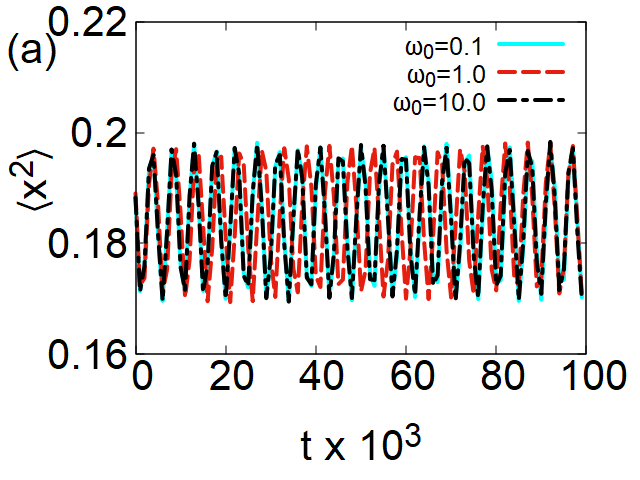} \hfil
\includegraphics[width=0.48\columnwidth]{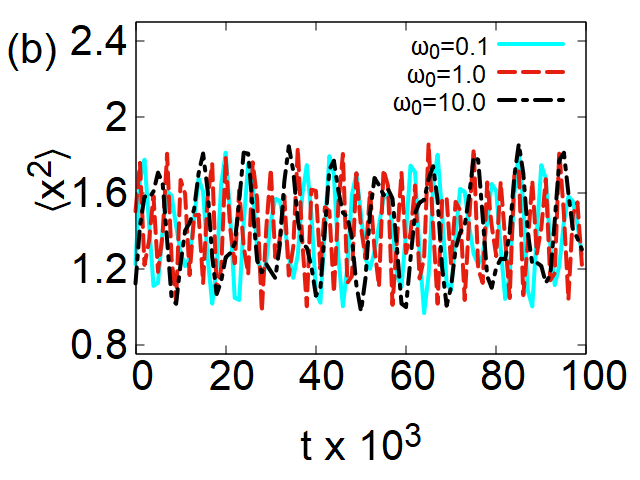} \caption{(a) Maximum amplitude of the perturbed numerical solution $|\psi|^{2}$
as a function of the coordinate $t$ for the case with a seesaw potential
(\ref{linear}), using the solutions constructed with the variational
parameters $A_{-}$ and $B_{-}$. (b) Average width of the numerical
solution as a function of $t$. The other parameter values used were
$E=-0.1$, $G=1$, $\Gamma=0.1$, and $b=2.0$. The three curves correspond
to different values of $\omega_{0}$, as indicated in the legend description.}
\label{F7}
\end{figure}

Now, using the modulation that leads to the seesaw-type potential,
given by Eq. (\ref{linear}), we performed new simulations considering
different values of $\omega_{0}$. As shown in Fig. \ref{F7}, for
values within the range $0.1\leq\omega_{0}\leq10$, all solutions
remain stable under the small perturbations introduced. It is worth
noting that although $h_{\max}$ and $\langle x^{2}\rangle$ exhibit
oscillations, these remain small and centered around a fixed value.
We also emphasize that, despite restricting Fig. \ref{F7} to only
three values of $\omega_{0}$, we conducted simulations for various
values within the range $[0.1,10]$, consistently verifying the stability
condition.

Next, we consider the solutions associated with the \emph{flying-bird}
potential (given by Eq. (\ref{fly})) for two different values of
$\sigma$, corresponding to distinct modulation amplitudes of these
solutions.

\begin{figure}[tb]
\centering \includegraphics[width=0.48\columnwidth]{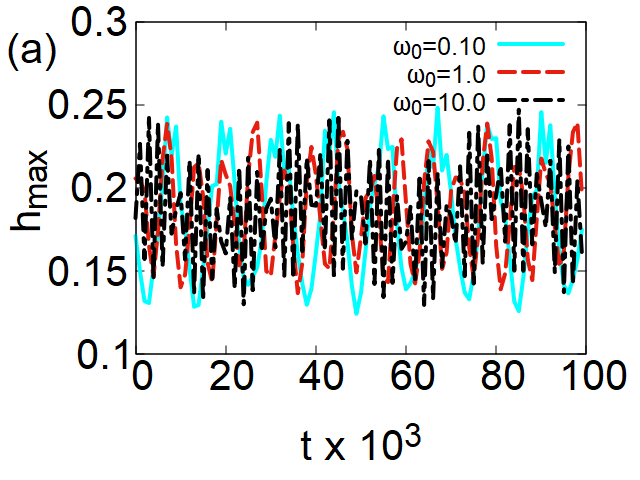} \hfil
\includegraphics[width=0.48\columnwidth]{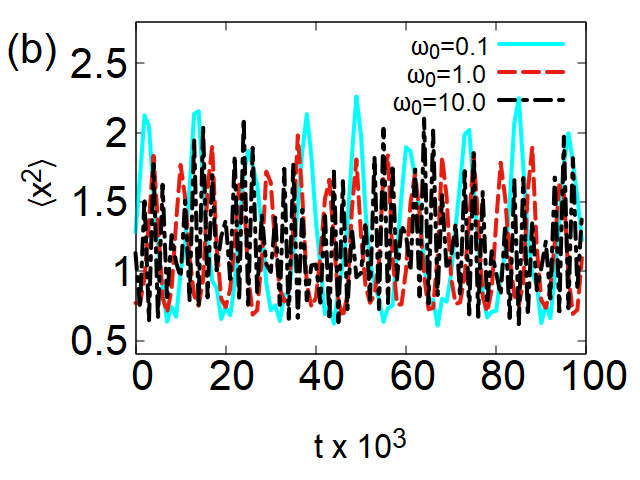} \caption{(a) Maximum amplitude of the perturbed numerical solution $|\psi|^{2}$
as a function of the coordinate $t$ for the case with a \emph{flying-bird}
potential (\ref{fly}), using the solutions constructed with the variational
parameters $A_{-}$ and $B_{-}$. (b) Average width of the numerical
solution as a function of $t$. The other parameter values used were
$E=-0.1$, $G=1$, $\Gamma=0.1$, $b=2.0$, and $\sigma=0.25$. The
three curves correspond to different values of $\omega_{0}$, as indicated
in the legend description.}
\label{F8}
\end{figure}

First, in Fig. \ref{F8}, we consider $\sigma=0.25$ and present the
results for three different values of $\omega_{0}$. It is observed
that for $\sigma=0.25$, the solutions remain stable for all values
of $\omega_{0}$. Note that the oscillations are now more pronounced
due to the amplitude modulation introduced by the $\alpha(t)$ term.
However, the mean values of both the amplitude and the solution width
remain constant, confirming the stability of the solutions. Furthermore,
for $\sigma=0.25$, stability was also observed in several additional
simulations conducted within the range $\omega_{0}=[0.1,10]$.

\begin{figure}[tb]
\centering \includegraphics[width=0.48\columnwidth]{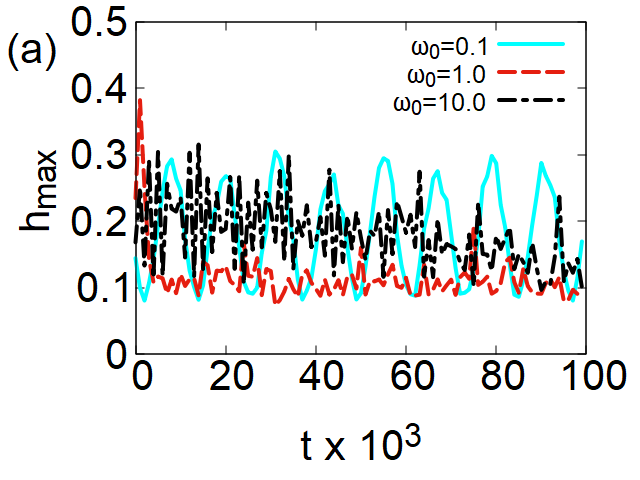} \hfil
\includegraphics[width=0.48\columnwidth]{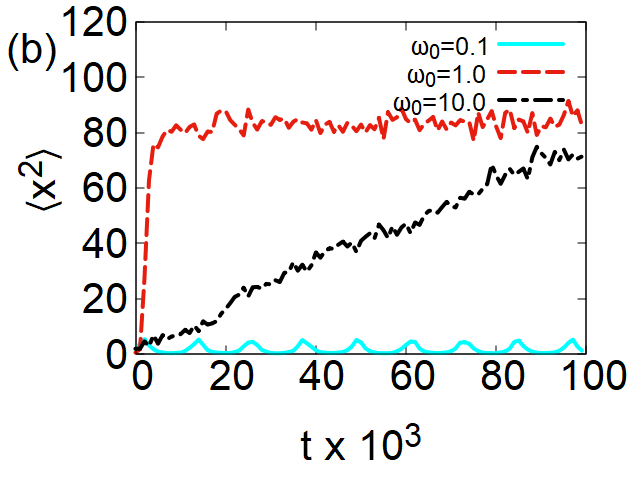} \hfil \includegraphics[width=0.8\columnwidth]{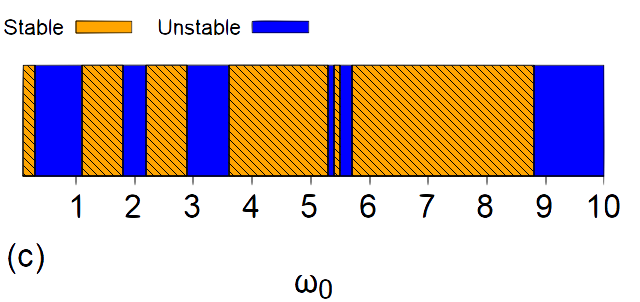}
\caption{(a) Maximum amplitude of the perturbed numerical solution $|\psi|^{2}$
as a function of the coordinate $t$ for the case with a \emph{flying-bird}
potential (\ref{fly}), using the solutions constructed with the variational
parameters $A_{-}$ and $B_{-}$. (b) Average width of the numerical
solution as a function of $t$. The other parameter values used were
$E=-0.1$, $G=1$, $\Gamma=0.1$, $b=2.0$, and $\sigma=0.5$. The
three curves correspond to different values of $\omega_{0}$, as indicated
in the legend description. (c) Stability/instability region as a function
of $\omega_{0}$.}
\label{F9}
\end{figure}

On the other hand, when considering $\sigma=0.5$ in $\alpha(t)$,
we observe a completely different behavior compared to the case with
$\sigma=0.25$. In this case, we identify multiple regions of $\omega_{0}$
(within $0.1\leq\omega_{0}\leq10$) where the solutions transition
between stability and instability. Panels (a) and (b) of Fig. \ref{F9}
display the amplitude and width of three solutions obtained for different
values of $\omega_{0}$. In Fig. \ref{F9}(c), we present the stability/instability
regions of the solutions as a function of $\omega_{0}$. Among the
three values of $\omega_{0}$ considered in this figure, only the
case $\omega_{0}=0.1$ remains stable.

By analyzing the stability transition points in Fig. \ref{F9}(c),
we find that the solutions remain stable within the shaded regions
of $\omega_{0}$ and become unstable in the solid regions. Overall,
we identify six stability regions and six distinct instability regions.
This result highlights the extreme sensitivity of the solutions to
the modulation being considered.

\begin{figure}[tb]
\centering \includegraphics[width=0.48\columnwidth]{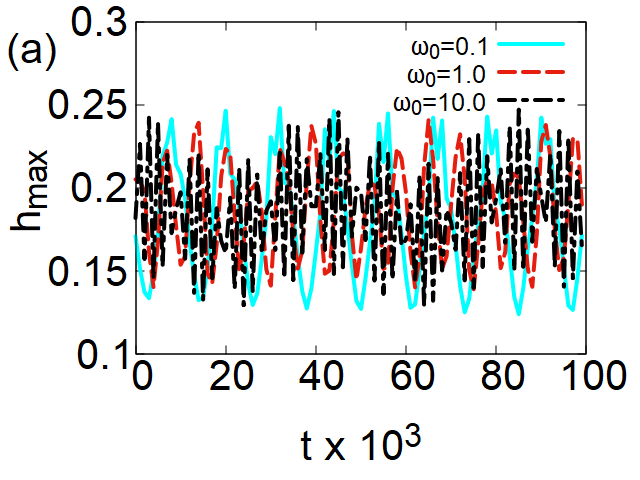} \hfil
\includegraphics[width=0.48\columnwidth]{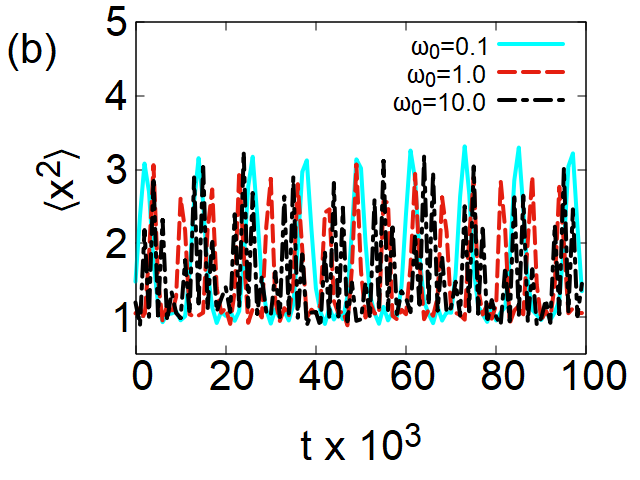} \caption{(a) Maximum amplitude of the perturbed numerical solution $|\psi|^{2}$
as a function of the coordinate $t$ for the case with the\emph{ mixed}
potential (\ref{mixed}), using the solutions constructed with the
variational parameters $A_{-}$ and $B_{-}$. (b) Average width of
the numerical solution as a function of $t$. The other parameter
values used were $E=-0.1$, $G=1$, $\Gamma=0.1$, $b=2.0$, and $\sigma=0.25$.
The three curves correspond to different values of $\omega_{0}$,
as indicated in the legend description.}
\label{F10}
\end{figure}

Finally, we analyze the solutions modulated by the mixed potential,
given by Eq. (\ref{mixed}). Similar to the previous case, when $\sigma=0.25$,
all the examined solutions remained stable within the range $0.1\leq\omega_{0}\leq10$.
As an example, in Fig. \ref{F10}, we present the results for three
values of $\omega_{0}$ that span the entire considered range.

\begin{figure}[tb]
\centering \includegraphics[width=0.48\columnwidth]{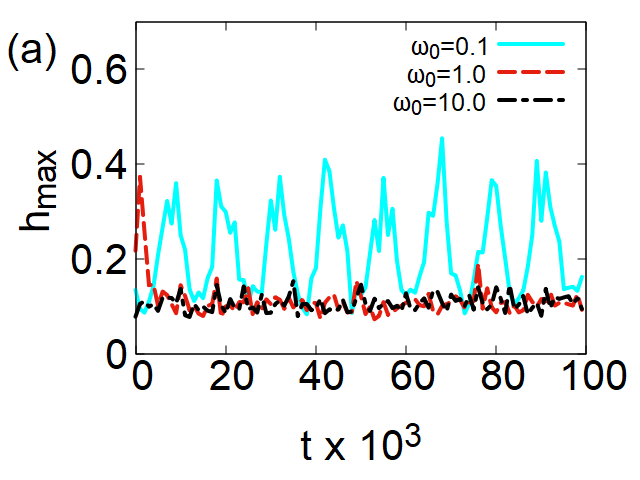}\hfil
\includegraphics[width=0.48\columnwidth]{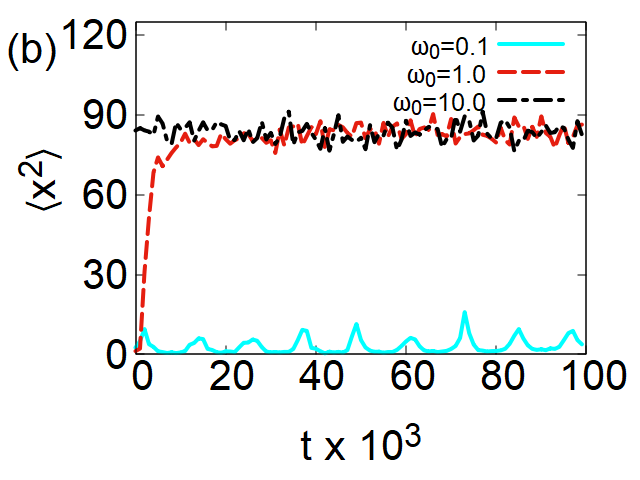} \hfil \includegraphics[width=0.8\columnwidth]{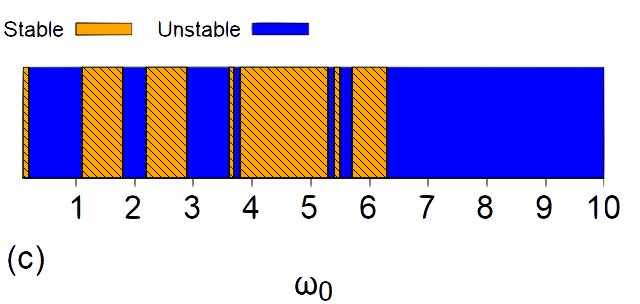}
\caption{(a) Maximum amplitude of the perturbed numerical solution $|\psi|^{2}$
as a function of the coordinate $t$ for the case with the \emph{mixed}
potential (\ref{mixed}), using the solutions constructed with the
variational parameters $A_{-}$ and $B_{-}$. (b) Average width of
the numerical solution as a function of $t$. The other parameter
values used were $E=-0.1$, $G=1$, $\Gamma=0.1$, $b=2.0$, and $\sigma=0.5$.
The three curves correspond to different values of $\omega_{0}$,
as indicated in the legend description. (c) Stability/instability
region as a function of $\omega_{0}$.}
\label{F11}
\end{figure}

On the other hand, when considering $\sigma=0.5$ in the modulation
induced by the mixed potential, we observe the existence of regions
where the solutions become unstable. Following the analysis pattern
from the previous case, panels (a) and (b) of Fig. \ref{F11} display
the amplitude and width of three solutions obtained for different
values of $\omega_{0}$. Once again, among the three considered values
of $\omega_{0}$, only the case $\omega_{0}=0.1$ remains stable.
This becomes evident when examining Fig. \ref{F11}(b), which shows
the abrupt growth of the solution width for $\omega_{0}=1$ and $\omega_{0}=10$.
However, as observed in Fig. \ref{F11}(c), the stability and instability
regions alternate throughout the range $0.1\leq\omega_{0}\leq10$.

Overall, we identified six stability regions and seven distinct instability
regions. Since this model corresponds to a linear combination of the
two previous models, one can observe that the instability arising
in the case $\sigma=0.5$ for the modulation induced by the mixed
potential is due to the inclusion of the \emph{flying-bird}-type term.
However, the stability regions underwent significant changes compared
to those presented in Fig. \ref{F9}(c).

\section{Conclusion}

In conclusion, this study explores the stability and dynamics of modulated
solutions in saturable nonlinear systems through both analytical and
numerical approaches. We analyzed various modulation patterns, namely
\emph{static}, \emph{seesaw}, \emph{flying-bird}, and \emph{mixed}
potentials, to investigate the impact on the stability of the solutions.
Our findings reveal that the stability of the solutions is highly
sensitive to the modulation parameters, with certain regions exhibiting
stable behavior while others lead to instability. Notably, the inclusion
of the \emph{flying-bird}-type term in the \emph{mixed} potential
significantly influences the stability characteristics. Overall, the
results underscore the importance of modulation in controlling the
stability of localized solutions in nonlinear systems, providing valuable
insights for future research in both theoretical and applied contexts.

\begin{acknowledgments}
We acknowledge the financial support provided by the Brazilian agencies
CNPq (grant \#306105/2022-5 and Sisphoton Laboratory- MCTI No. 440225/2021-3),
CAPES, and FAPEG. This work was also performed as part of the Brazilian
National Institute of Science and Technology (INCT) for Quantum Information
(Grant No. 465469/2014-0).
\end{acknowledgments}

\bibliographystyle{apsrev4-2}
\bibliography{Refs}

@book{Yang_10,
author = {Yang, Jianke},
doi = {10.1137/1.9780898719680},
isbn = {978-0-89871-705-1},
month = {jan},
publisher = {Society for Industrial and Applied Mathematics},
title = {{Nonlinear Waves in Integrable and Nonintegrable Systems}},
url = {http://epubs.siam.org/doi/book/10.1137/1.9780898719680},
year = {2010}
}

@article{Yan_PLA10,
author = {Yan, Zhenya},
doi = {10.1016/j.physleta.2009.11.030},
issn = {03759601},
journal = {Phys. Lett. A},
month = {jan},
number = {4},
pages = {672--679},
title = {{Nonautonomous "rogons" in the inhomogeneous nonlinear Schr{\"{o}}dinger equation with variable coefficients}},
url = {http://linkinghub.elsevier.com/retrieve/pii/S0375960109014625},
volume = {374},
year = {2010}
}

@article{Calaca_CNSNS14,
author = {Cala{\c{c}}a, L and Avelar, A.T. and Bazeia, D and Cardoso, W.B.},
doi = {10.1016/j.cnsns.2014.02.002},
issn = {10075704},
journal = {Commun. Nonlinear Sci. Numer. Simul.},
month = {sep},
number = {9},
pages = {2928--2934},
title = {{Modulation of localized solutions for the Schr{\"{o}}dinger equation with logarithm nonlinearity}},
url = {http://linkinghub.elsevier.com/retrieve/pii/S1007570414000550},
volume = {19},
year = {2014}
}

@article{Belmonte-Beitia_PRL08,
author = {Belmonte-Beitia, Juan and P{\'{e}}rez-Garc{\'{i}}a, V{\'{i}}ctor M. and Vekslerchik, Vadym and Konotop, Vladimir V.},
doi = {10.1103/PhysRevLett.100.164102},
issn = {0031-9007},
journal = {Phys. Rev. Lett.},
month = {apr},
number = {16},
pages = {164102},
title = {{Localized Nonlinear Waves in Systems with Time- and Space-Modulated Nonlinearities}},
url = {http://link.aps.org/doi/10.1103/PhysRevLett.100.164102},
volume = {100},
year = {2008}
}

@book{Agrawal_13,
author = {Agrawal, G P},
isbn = {9780123970237},
publisher = {Elsevier Science},
series = {Optics and Photonics},
title = {{Nonlinear Fiber Optics}},
url = {https://books.google.com.br/books?id=xNvw-GDVn84C https://books.google.com.br/books?id=b5S0JqHMoxAC https://books.google.com.br/books?id=wjHP0oAVcScC},
year = {2013}
}

@article{Cardoso_CNSNS17,
author = {Cardoso, Wesley B. and Couto, Hugo L.C. and Avelar, Ardiley T. and Bazeia, Dionisio},
doi = {10.1016/j.cnsns.2017.01.012},
issn = {10075704},
journal = {Commun. Nonlinear Sci. Numer. Simul.},
month = {jul},
pages = {474--483},
title = {{Modulation of localized solutions in quadratic-cubic nonlinear Schr{\"{o}}dinger equation with inhomogeneous coefficients}},
url = {http://linkinghub.elsevier.com/retrieve/pii/S1007570417300114},
volume = {48},
year = {2017}
}

@article{Saravanan_CNSNS19,
author = {Saravanan, M. and Cardoso, Wesley B.},
doi = {10.1016/j.cnsns.2018.09.021},
issn = {10075704},
journal = {Commun. Nonlinear Sci. Numer. Simul.},
month = {apr},
pages = {176--186},
title = {{Parametrically driven localized magnetic excitations with spatial inhomogeneity}},
url = {https://linkinghub.elsevier.com/retrieve/pii/S1007570418303010},
volume = {69},
year = {2019}
}

@article{Cardoso_NA10,
author = {Cardoso, W.B. and Avelar, A.T. and Bazeia, D.},
doi = {10.1016/j.nonrwa.2010.05.013},
issn = {14681218},
journal = {Nonlinear Anal. Real World Appl.},
month = {oct},
number = {5},
pages = {4269--4274},
title = {{Bright and dark solitons in a periodically attractive and expulsive potential with nonlinearities modulated in space and time}},
url = {http://linkinghub.elsevier.com/retrieve/pii/S1468121810000805},
volume = {11},
year = {2010}
}

@article{Avelar_PRE09,
archivePrefix = {arXiv},
arxivId = {arXiv:0902.3135v1},
author = {Avelar, A. T. and Bazeia, D. and Cardoso, W. B.},
doi = {10.1103/PhysRevE.79.025602},
eprint = {arXiv:0902.3135v1},
isbn = {1539-3755},
issn = {1539-3755},
journal = {Phys. Rev. E},
month = {feb},
number = {2},
pages = {025602},
title = {{Solitons with cubic and quintic nonlinearities modulated in space and time}},
url = {http://link.aps.org/doi/10.1103/PhysRevE.79.025602 https://link.aps.org/doi/10.1103/PhysRevE.79.025602},
volume = {79},
year = {2009}
}

@article{Cardoso_PRE12,
author = {Cardoso, W. B. and Avelar, A. T. and Bazeia, D.},
doi = {10.1103/PhysRevE.86.027601},
issn = {1539-3755},
journal = {Phys. Rev. E},
month = {aug},
number = {2},
pages = {027601},
publisher = {American Physical Society},
title = {{Modulation of localized solutions in a system of two coupled nonlinear Schr{\"{o}}dinger equations}},
url = {http://link.aps.org/doi/10.1103/PhysRevE.86.027601 https://link.aps.org/doi/10.1103/PhysRevE.86.027601},
volume = {86},
year = {2012}
}

@article{Cardoso_PRE13,
author = {Cardoso, W. B. and Zeng, J. and Avelar, A. T. and Bazeia, D. and Malomed, B. A.},
doi = {10.1103/PhysRevE.88.025201},
issn = {1539-3755},
journal = {Phys. Rev. E},
month = {aug},
number = {2},
pages = {025201},
publisher = {American Physical Society},
title = {{Bright solitons from the nonpolynomial Schr{\"{o}}dinger equation with inhomogeneous defocusing nonlinearities}},
url = {http://link.aps.org/doi/10.1103/PhysRevE.88.025201 https://link.aps.org/doi/10.1103/PhysRevE.88.025201},
volume = {88},
year = {2013}
}

@article{Salasnich_PRA14,
author = {Salasnich, Luca and Cardoso, Wesley B. and Malomed, Boris A.},
doi = {10.1103/PhysRevA.90.033629},
issn = {1050-2947},
journal = {Phys. Rev. A},
month = {sep},
number = {3},
pages = {033629},
title = {{Localized modes in quasi-two-dimensional Bose-Einstein condensates with spin-orbit and Rabi couplings}},
url = {https://link.aps.org/doi/10.1103/PhysRevA.90.033629},
volume = {90},
year = {2014}
}

@article{Avelar_PRE10,
author = {Avelar, A.T. T. and Bazeia, D. and Cardoso, W.B. B.},
doi = {10.1103/PhysRevE.82.057601},
issn = {1539-3755},
journal = {Phys. Rev. E},
month = {nov},
number = {5},
pages = {057601},
title = {{Modulation of breathers in the three-dimensional nonlinear Gross-Pitaevskii equation}},
url = {https://link.aps.org/doi/10.1103/PhysRevE.82.057601 http://link.aps.org/doi/10.1103/PhysRevE.82.057601},
volume = {82},
year = {2010}
}

@article{Santos_ND22,
author = {dos Santos, Renato D. and Cardoso, Wesley B.},
doi = {10.1007/s11071-021-07090-y},
issn = {0924-090X},
journal = {Nonlinear Dyn.},
month = {jan},
number = {1},
pages = {1205--1214},
title = {{Modulation of localized solutions of an inhomogeneous cigar-shaped superfluid fermion gas}},
url = {https://link.springer.com/10.1007/s11071-021-07090-y},
volume = {107},
year = {2022}
}

@article{Cardoso_BJP21,
author = {Cardoso, Wesley Bueno and Avelar, Ardiley Torres and Bazeia, Dionisio},
doi = {10.1007/s13538-020-00836-w},
issn = {0103-9733},
journal = {Brazilian J. Phys.},
month = {apr},
number = {2},
pages = {151--156},
title = {{Propagation of Solitons in Quasi-periodic Nonlinear Coupled Waveguides}},
url = {http://link.springer.com/10.1007/s13538-020-00836-w},
volume = {51},
year = {2021}
}

@article{Calaca_OQE17,
author = {Cala{\c{c}}a, Luciano and Cardoso, Wesley B.},
doi = {10.1007/s11082-017-1214-1},
issn = {0306-8919},
journal = {Opt. Quantum Electron.},
month = {nov},
number = {11},
pages = {379},
title = {{Modulation of localized solutions in an inhomogeneous saturable nonlinear Schr{\"{o}}dinger equation}},
url = {http://link.springer.com/10.1007/s11082-017-1214-1},
volume = {49},
year = {2017}
}

@article{Cardoso_PLA10-2,
author = {Cardoso, W. B. and Avelar, A. T. and Bazeia, D. and Hussein, M. S.},
doi = {10.1016/j.physleta.2010.03.065},
issn = {03759601},
journal = {Phys. Lett. A},
month = {may},
number = {23},
pages = {2356--2360},
title = {{Solitons of two-component Bose--Einstein condensates modulated in space and time}},
url = {https://linkinghub.elsevier.com/retrieve/pii/S0375960110004068},
volume = {374},
year = {2010}
}

@article{Borovkova_PRE11,
author = {Borovkova, Olga V and Kartashov, Yaroslav V and Torner, Lluis and Malomed, Boris A},
doi = {10.1103/PhysRevE.84.035602},
issn = {1539-3755},
journal = {Phys. Rev. E},
month = {sep},
number = {3},
pages = {35602},
title = {{Bright solitons from defocusing nonlinearities}},
url = {https://link.aps.org/doi/10.1103/PhysRevE.84.035602},
volume = {84},
year = {2011}
}

@book{Pethick_08,
author = {Pethick, C. J. and Smith, H.},
doi = {10.1017/CBO9780511802850},
isbn = {9780521846516},
month = {sep},
publisher = {Cambridge University Press},
title = {{Bose--Einstein Condensation in Dilute Gases}},
url = {https://www.cambridge.org/core/product/identifier/9780511802850/type/book},
year = {2008}
}

@article{Cardoso_PLA10,
author = {Cardoso, W. B. and Avelar, A. T. and Bazeia, D.},
doi = {10.1016/j.physleta.2010.04.050},
issn = {03759601},
journal = {Phys. Lett. A},
month = {jun},
number = {26},
pages = {2640--2645},
title = {{Modulation of breathers in cigar-shaped Bose--Einstein condensates}},
url = {http://linkinghub.elsevier.com/retrieve/pii/S0375960110004895 https://linkinghub.elsevier.com/retrieve/pii/S0375960110004895},
volume = {374},
year = {2010}
}

@book{Kivshar_03,
author = {Kivshar, Yuri S. and Agrawal, Govind P.},
doi = {10.1016/B978-0-12-410590-4.X5000-1},
isbn = {9780124105904},
publisher = {Elsevier},
title = {{Optical Solitons: From Fibers to Photonic Crystals}},
url = {https://linkinghub.elsevier.com/retrieve/pii/B9780124105904X50001},
year = {2003}
}

@book{Pitaevskii_16,
author = {Pitaevskii, Lev and Stringari, Sandro},
doi = {10.1093/acprof:oso/9780198758884.001.0001},
isbn = {019875888X},
month = {jan},
publisher = {Oxford University PressOxford},
title = {{Bose-Einstein Condensation and Superfluidity}},
url = {https://academic.oup.com/book/26678},
year = {2016}
}

@book{Sulem_04,
address = {New York, NY},
author = {Sulem, Catherine and Sulem, Pirre-Louis},
doi = {10.1007/b98958},
editor = {Sulem, Catherine and Sulem, Pirre-Louis},
isbn = {978-0-387-98611-1},
publisher = {Springer New York},
series = {Applied Mathematical Sciences},
title = {{The Nonlinear Schr{\"{o}}dinger Equation: Self-Focusing and Wave Collapse}},
url = {http://link.springer.com/10.1007/b98958},
volume = {139},
year = {2004}
}

@article{Maddouri_PLA24,
author = {Maddouri, Kamel and Triki, Houria and Wang, Nan and Zhou, Qin},
doi = {10.1016/j.physleta.2024.129584},
issn = {03759601},
journal = {Phys. Lett. A},
month = {jul},
pages = {129584},
title = {{Nonlinear tunneling of chirped similaritons in non-centrosymmetric waveguides with quadratic-cubic nonlinearity}},
url = {https://linkinghub.elsevier.com/retrieve/pii/S0375960124002780},
volume = {512},
year = {2024}
}

@article{Uthayakumar_FP20,
author = {Uthayakumar, T. and {Al Sakkaf}, L. and {Al Khawaja}, U.},
doi = {10.3389/fphy.2020.596886},
issn = {2296-424X},
journal = {Front. Phys.},
month = {dec},
title = {{Peregrine Solitons of the Higher-Order, Inhomogeneous, Coupled, Discrete, and Nonlocal Nonlinear Schr{\"{o}}dinger Equations}},
url = {https://www.frontiersin.org/articles/10.3389/fphy.2020.596886/full},
volume = {8},
year = {2020}
}

@article{Oztas_PLA24,
author = {Oztas, Z. and Kaplan, E.},
doi = {10.1016/j.physleta.2024.129853},
issn = {03759601},
journal = {Phys. Lett. A},
month = {nov},
pages = {129853},
title = {{Exact soliton solutions of Gross Pitaevskii equation with a variable shape optical lattice potential}},
url = {https://linkinghub.elsevier.com/retrieve/pii/S0375960124005474},
volume = {525},
year = {2024}
}

@article{Miranda_OQE24,
author = {Miranda, Bruno M. and Avelar, Ardiley T. and Cardoso, Wesley B. and Bazeia, Dionisio},
doi = {10.1007/s11082-024-07757-x},
issn = {1572-817X},
journal = {Opt. Quantum Electron.},
month = {nov},
number = {12},
pages = {1915},
title = {{Dynamics of localized solutions in three core coupled waveguides with quasi-periodic nonlinearity}},
url = {https://link.springer.com/10.1007/s11082-024-07757-x},
volume = {56},
year = {2024}
}

@article{Cardoso_ND21,
author = {Cardoso, W. B. and Avelar, A. T. and Bazeia, D.},
doi = {10.1007/s11071-021-06962-7},
issn = {0924-090X},
journal = {Nonlinear Dyn.},
month = {dec},
number = {4},
pages = {3469--3477},
title = {{Effects of chaotic perturbations on a nonlinear system undergoing two-soliton collisions}},
url = {https://link.springer.com/10.1007/s11071-021-06962-7},
volume = {106},
year = {2021}
}

@article{Calaca_EPJST18,
author = {Cala{\c{c}}a, Luciano and Avelar, Ardiley T. and Malomed, Boris A. and Cardoso, Wesley B.},
doi = {10.1140/epjst/e2018-00118-5},
issn = {1951-6355},
journal = {Eur. Phys. J. Spec. Top.},
month = {sep},
number = {5-6},
pages = {551--561},
title = {{Influence of pseudo-stimulated-Raman-scattering on the modulational instability in an inhomogeneous nonlinear medium}},
url = {http://link.springer.com/10.1140/epjst/e2018-00118-5},
volume = {227},
year = {2018}
}

@article{Nath_EPJD22,
author = {Nath, Ajay and Bera, Jayanta and Pathak, Maitri R. and Roy, Utpal},
doi = {10.1140/epjd/s10053-022-00571-8},
issn = {1434-6060},
journal = {Eur. Phys. J. D},
month = {dec},
number = {12},
pages = {241},
title = {{Solitary matter waves in a tunable bi-periodic optical lattice with two- and three-body interactions}},
url = {https://link.springer.com/10.1140/epjd/s10053-022-00571-8},
volume = {76},
year = {2022}
}

@article{Belmonte-Beitia_JPA09,
author = {Belmonte-Beitia, J and Cuevas, J},
doi = {10.1088/1751-8113/42/16/165201},
issn = {1751-8113},
journal = {J. Phys. A Math. Theor.},
month = {apr},
number = {16},
pages = {165201},
title = {{Solitons for the cubic-quintic nonlinear Schr{\"{o}}dinger equation with time- and space-modulated coefficients}},
url = {https://iopscience.iop.org/article/10.1088/1751-8113/42/16/165201},
volume = {42},
year = {2009}
}

@article{Rocha_ND23,
author = {da Rocha, Maurilho R and Avelar, Ardiley T and Cardoso, Wesley B},
doi = {10.1007/s11071-022-08104-z},
issn = {0924-090X},
journal = {Nonlinear Dyn.},
month = {mar},
number = {5},
pages = {4769--4777},
title = {{Localized solutions of inhomogeneous saturable nonlinear Schr{\"{o}}dinger equation}},
url = {https://link.springer.com/10.1007/s11071-022-08104-z},
volume = {111},
year = {2023}
}

\end{document}